\documentclass[preprint]{imsart}

\RequirePackage[OT1]{fontenc}
\RequirePackage{amsthm,amsmath}
\RequirePackage[numbers]{natbib}
\RequirePackage[colorlinks,citecolor=blue,urlcolor=blue]{hyperref}

\usepackage{figsize,subfigure,ifthen,calc,psfrag,bm}
\usepackage{enumerate,comment}
\usepackage{xspace}
\usepackage{latexsym}
\usepackage{amsfonts}
\usepackage{rotating}
\allowdisplaybreaks
\usepackage{amssymb}
\usepackage{dsfont}
\usepackage{nameref}
\startlocaldefs
\numberwithin{equation}{section}
\theoremstyle{plain}

\newcommand\be{\begin{eqnarray}}
\newcommand\ee{\end{eqnarray}}
\newcommand\bee{\begin{eqnarray*}}
\newcommand\eee{\end{eqnarray*}}

\newtheorem{lemma}{Lemma}[section]
\newtheorem{theorem}[lemma]{Theorem}

\newtheorem{example}[lemma]{Example}

\theoremstyle{definition}
\newtheorem{remark}[lemma]{Remark}

\theoremstyle{plain}
\newcounter{ASSUMPTION}
\newtheorem{assumption}[ASSUMPTION]{Assumption}

\newcommand{\f}{{\bf f}}

\def\C{{\mathbb C}}
\def\R {{\mathbb R}}

\def\N{{\mathbb N}}
\def\Z{{\mathbb Z}}

\def\UMa{{\bf Id}}

\endlocaldefs

\begin{document}

\begin{frontmatter}
\title{A Frequency Domain Bootstrap for General Multivariate Stationary Processes%\thanksref{T1}
}
\runtitle{Multivariate Frequency Domain  Bootstrap}
%\thankstext{T1}{Footnote to the title with the ``thankstext'' command.}

\begin{aug}
\author{\fnms{Marco} \snm{Meyer}\thanksref{}\ead[label=e1]{marco.meyer@tu-bs.de}}
\and
\author{\fnms{Efstathios} \snm{Paparoditis}\thanksref{t2}
\ead[label=e2]{stathisp@ucy.ac.cy}}
%\and
%\author{\fnms{Jens-Peter} \snm{Kreiss}\thanksref{t1}
%\ead[label=e3]{j.kreiss@tu-bs.de}
%\ead[label=u1,url]{stathisp@ucy.a.c.cy}
%}

%\thankstext{t1}{Some comment}
\thankstext{t2}{Supported in part by a University of Cyprus Research Grant.}
%\thankstext{t3}{Second supporter of the project}
\runauthor{M. Meyer and E. Paparoditis}

\affiliation{Technische Universit\"at Braunschweig\thanksmark{t1} and University of Cyprus\thanksmark{t2}}

\address{Technische Universit\"at Braunschweig\\ Inst. f. Math. Stochastik\\ Universit\"atsplatz 2\\ D--38106 Braunschweig, Germany.\\
\printead{e1}\\
%\phantom{E-mail:\ }\printead*{e3}
}

\address{University of Cyprus \\Department of Mathematics and Statistics\\
P.O.Box 20537, CY-1678 Nicosia\\ Cyprus.\\
\printead{e2}\\
%\printead{u1}
%\phantom{E-mail:\ }\printead*{e2}
}
\end{aug}

\begin{abstract}
For many relevant statistics of multivariate time series, no valid frequency domain bootstrap procedures exist. This is mainly due to the fact that the distribution of such statistics depends on the fourth-order moment structure of the underlying process in nearly every scenario, except for  some special cases  like Gaussian time series.
%Many of the established bootstrap procedures  for univariate time series, both in time and in frequency domain,
% are in general not able to imitate these fourth-order quantities in the multivariate context.
In contrast to the univariate case, even additional structural assumptions such as linearity of the multivariate process or a standardization of the statistic of interest do not solve the problem.
% in the multivariate case.
This paper  focuses on integrated periodogram statistics  as well as functions thereof and presents a new frequency domain bootstrap  procedure for multivariate time series, the multivariate frequency domain hybrid bootstrap  (MFHB), to fill this gap.
%The procedure combines two ingredients: The first uses the asymptotic independence and the complex Wishart distribution of the periodogram matrix of a  general stationary process in order
%to imitate  those  features of the target distribution that depend on the second-order structure of the process. The second ingredient, which is
%an adaptation of the idea of convolved subsampling in the multivariate frequency domain context,  imitates   all  desired high-order features of this distribution beyond those of order two.
%Since the statistics under consideration are in the multivariate context complex-valued, the MFB has to  establish the appropriate fourth-order moment structure of both, the covariance and the relation matrix simultaneously.
Asymptotic validity  of the MFHB procedure is established  for general classes of periodogram-based statistics and for  stationary  multivariate processes satisfying rather weak dependence conditions. A simulation study is carried out which compares the finite sample performance of the MFHB with that of the  moving block bootstrap.
% ranging clearly far beyond linear processes.
\end{abstract}



\end{frontmatter}

\section{Introduction}

Developing valid bootstrap methods for
%multivariate
time series has been a particularly challenging problem since the mid-1980s. Soon after Efron's seminal paper (cf. Efron (1979)) on the bootstrap for i.i.d.~observations the first attempts towards an extension to dependent data were made. While for univariate time series a variety of proposals exists that are asymptotically valid under certain assumptions on the dependence structure of the underlying process and for certain types of statistics, very little progress has been made for multivariate time series. This is due to  the fact  that in the multivariate context the distribution of most relevant statistics depends on the fourth-order moment structure of the underlying process, which many bootstrap methods  developed for univariate time series are not able to imitate. Even under the assumption of a linear time series, that is, an $ \R^m $-valued, strictly stationary stochastic process $ ({\bf X}(t))_{t \in \Z} $ which fullfils
\be
{\bf X}(t) = \sum_{j=-\infty}^{\infty} {\bf B}_j \, {\bf e}(t-j), \ \  t\in \Z, \label{linearproc}
\ee
for certain $ m \times m $ real matrices $ {\bf B}_j $ and an i.i.d.~white noise process $ ({\bf e}(t))_{t \in \Z} $, the distribution of most statistics of interest depends on the fourth-order moment structure and many bootstrap methods fail.
%We mention  here  sample crosscorrelations as an important statistic in this context.
 In contrast to this, for univariate  linear time series --  that is for processes \eqref{linearproc} with dimension $ m=1 $ -- there are a number of scenarios in which the distribution of  some statistic of interest depends only on first and second-order moments
% -- particularly for linear time series such as \eqref{linearproc} with dimension $ m=1 $ --
and established univariate bootstrap methods are successful.
%This is for instance true for sample autocorrelations. We refer here to  Dahlhaus and Janas (1996)  for frequency domain and to Kreiss et al. (2011) for time domain bootstrap procedures; see also Kreiss and Paparoditis (20XX) for an overview.\\
It should be emphasized at this point that whenever we use the term \emph{linear time series (or linear process)} in this work, we refer to a process as given by \eqref{linearproc} \emph{including} the i.i.d.~assumption on the innovations ${\bf e}(t)$, as it is done in many standard references for time series, cf. Brockwell and Davis (1991), among others.  A subclass of (\ref{linearproc})  is that of  the causal and invertible linear processes. In this case a sequence  $ \{{\bf A}_j, j \in \N\}$  of $m\times m$ real matrices exists with
$ \sum_{j=1}^\infty \|{\bf A}_j\|_F <\infty$, such that
(\ref{linearproc})  also can be expressed as  ${\bf X}(t) = \sum_{j=1}^{\infty} {\bf A}_j \, {\bf X}(t-j) + {\bf e}(t)$.   $ \{{\bf X}(t),t \in \Z\}$   is   then called a linear VAR($\infty$) process.  Here $ \|\cdot\|_F$ refers to  the Frobenius norm of a matrix.  Notice that  any process following   expression (\ref{linearproc}) but with non-i.i.d.~white noise innovations, i.e.~where the $ {\bf e}_t$ are uncorrelated with zero mean but not independent, is nonlinear.
%This notion is particularly important in the ensuing discussion on the range of validity of established bootstrap methods because most of the classical methods are valid exclusively for linear processes in this strict sense. To put this into perspective, processes with a Wold representation similar to \eqref{linearproc} but with only uncorrelated (not i.i.d.) innovations are considered nonlinear. The same goes for popular models such as, for instance, univariate bilinear processes and autoregressive, moving average or ARMA processes driven by ARCH noise or any other non-i.i.d~noise. Moreover, all linear processes are obviously strictly stationary which implies that any process which is not strictly stationary is nonlinear. \\

Bootstrap methods for time series are usually formulated either in the time domain or in the  frequency domain, with a few hybrid methods that combine both approaches. In the following paragraphs we will give a short overview of existing methods for the univariate and the multivariate setup, and discuss their respective limitations. 
Prominent examples in the time domain  are the block bootstrap and variants thereof, the AR sieve bootstrap and the linear process bootstrap, among others. Block bootstrap methods  are  rough tools which are valid for a wide class of processes  but typically  their performance heavily depends on the block size.
% and they exhibit
%rather slow convergence rates
%a rather inferior performance in simulation studies; see also Section 4.
For univariate time series,   the AR sieve bootstrap  is known to be valid exclusively in those situations where the distribution of interest depends on the first and second-order characteristics of the process only, cf. Kreiss, Paparoditis and Politis (2011).
There are some relevant statistics for which this is the case, like the sample mean  for general stationary processes or sample autocorrelations for linear processes while sample autocovariances (even for linear processes) are already outside the range of its validity.  The same remarks also can be made for the linear process bootstrap.   To overcome some of the aforementioned  limitations  for   univariate time series, Frangeskou and Paparoditis (2019) proposed a procedure that involves a wild bootstrap scheme to generate pseudo innovations which asymptotically correctly imitate the first, the second and the fourth-order moment structure of the true innovations. This  extends the validity
of the AR sieve bootstrap beyond the class of linear, causal and invertible processes.
%The basic idea  of a  corresponding  bootstrap procedure  is  to  use  a consistent, nonparametric   estimator of   the fourth-order cumulant of the i.i.d.~innovations  and to resample the pseudo errors   using a wild bootstrap  scheme. The generated  i.i.d.~pseudo innovations  imitate then asymptotically correct the first, the second {\it and} the fourth order moment structure of the true innovations.

However,  if one switches to multivariate processes the fourth-order moment structure  shows up in the asymptotic distribution  of almost any relevant statistic. This
%, including for instance sample cross-correlations. This
 makes the bootstrap estimation problem much more involved. In particular, the AR sieve and the linear process bootstrap fail  for  multivariate noninvertible  linear processes (\ref{linearproc}) and   for such basic statistics as sample cross-correlations; see Jentsch and Kreiss  (2010) and Meyer and Kreiss (2015) for more details.
  Thus, beyond causal and invertible linear processes, and apart from special cases like Gaussian time series, it seems that validity of the vector AR sieve bootstrap is essentially restricted to  very elementary statistics like the sample mean. Even the aforementioned extension of the AR sieve via wild bootstrap-generated pseudo innovations  is not available in the multivariate context.
% However,  for multivariate time series, it is not clear how such an approach can be implemented.
%This is due to the fact  that in contrast to the univariate case,  no consistent, nonparametric  estimators for the entire set of the fourth order cumulants  $ {\rm cum}(e_{i,t},e_{j,t}, e_{r,t},e_{s,t})$,  $ i,j,r,s \in \{1,2, \ldots, m\}$, of the $m$-dimensional i.i.d. innovations  in (\ref{linearproc})  exist. Furthermore, even if one succeeds in consistently estimating these quantities,   it is not clear how these estimators can be used in order to generate a sequence of $m$-dimensional i.i.d.~pseudo innovations  which  correctly imitate the first, the second and the entire fourth-order structure of the true vector of  i.i.d.~innovations ${\bf e}_t$. 
Consequently,   for a wide class of stationary multivariate processes, including linear processes (\ref{linearproc}),  and for many  interesting statistics,  the only available   bootstrap method is essentially the time domain block bootstrap and variants thereof.
%\footnote{MM: Is the the same for the linear process bootstrap? Also, more examples here, like AR-aided block BS?}\\

Concerning frequency domain bootstrap methods  for univariate time series, the situation is very similar.  Interest is here  focused on so-called integrated periodogram statistics. These are statistics which are obtained by integrating  over all frequencies the periodogram  multiplied  with some  function of interest.  Many time domain statistics like autocorrelations or autocovariances have a frequency domain analogue, that is they can also be expressed  as
 (functions of)  integrated periodograms. Hurvich and Zeger (1987), Franke and H\"ardle (1992) and Dahlhaus and Janas (1996) developed   for univariate time series a multiplicative bootstrap procedure  for the periodogram.
% for this class of integrated periodogram statistics.
By construction, this scheme is capable of imitating the variance of periodogram ordinates at different frequencies but not their  dependence structure across frequencies. Since for periodogram-based statistics,  the fourth-order structure of the process is transmitted  to their distribution  through the weak dependence of the periodogram ordinates across different frequencies,
the multiplicative periodogram  bootstrap suffers from the same limitations as some of the time domain procedures discussed so far. More specifically,   it is valid exclusively in those situations where the distribution of interest only depends on the spectral density, that is on the second-order structure of the process.  Dahlhaus and Janas (1996) showed that a subclass of standardized integrated periodograms, the so-called \emph{ratio statistics}, falls into this category, but only under the assumption that the underlying time series is linear. 
%For instance,  this  frequency domain approach works for sample autocorrelations (but not for sample autocovariances) of univariate linear time series, while the approach fails whenever the assumption of linearity in the strict sense (driven by i.i.d.~noise) is dropped.
% for integrated periodogram statistics.
%For univariate time series,
%a  procedure
 Kreiss and Paparoditis (2012) proposed  a modification of the multiplicative periodogram bootstrap which extends its  validity  for integrated periodograms  to the entire class of univariate linear  time series.
 %This approach uses an estimator of the fourth order cumulant of the i.i.d.  innovations of the underlying linear process in order to appropriately resample  the periodogram in a way which  mimics the weak dependence of periodogram ordinates across frequencies.
Recently, Meyer et al.  (2020)  proposed  a hybrid method 
%to bootstrap integrated periodogram statistics for univariate time series
which is valid
 for a very general class of  weakly dependent univariate processes and which goes far beyond the linear process class.

However, for multivariate time series the related   inference  problems are much more involved  and  no valid frequency domain bootstrap method exists so far.
%To elaborate, let the $ m$-variate time series of length $ n $ be denoted by $ {\bf X}(1),{\bf X}(2),\ldots,{\bf X}(n) $.
% To elaborate. many important statistics for multivariate time series can be written as functions of the periodogram matrix,
 % where the later  is defined
%at hand. Then the periodogram matrix of this series is defined
%Recall the definition of the periodogram matrix
%$ {\bf I}: [-\pi,\pi] \rightarrow \C^{m \times m} $ with $ {\bf I}(\lambda)={\bf d}(\lambda) \, \overline{\bf d}(\lambda)$, where $ \overline{\bf d}(\lambda) $ denotes the conjugate transpose of $ {\bf d}(\lambda) $, and the vector
%\bee
%{\bf d}(\lambda)= \frac{1}{\sqrt{2\pi n}} \sum_{t=1}^{n} {\bf X}(t) \, e^{-it\lambda}\,, \quad r=1,2, \ldots, m\,,
%\eee
%denotes the finite Fourier transform of the series at frequency $ \lambda \in [-\pi,\pi] $.
%To elaborate, recall first that important statistics such as sample cross-covariance and sample cross-correlation matrices can be written as integrated periodogram statistics  or as    functions thereof; see Section 2 for details.
In principle, an analogue of the univariate multiplicative approach (Franke and H\"ardle (1992) and  Dahlhaus and Janas (1996)),  can be formulated in the multivariate setup  by using the fact that -- for a wide class of stationary processes -- the periodogram matrices   at  any fixed set of frequencies
%$ 0 <\lambda_1 <\lambda_2 < \cdots < \lambda_m <\pi$,
are asymptotically independent and   have  a Wishart distribution;
%the  parameters of which depend on the spectral density matrix  of the process at
%the corresponding  frequencies;
see for instance Brillinger (1981).  However, applying such   a bootstrap approach alone will fail in imitating the dependence across periodogram ordinates   since it only  can capture  the second-order properties of the underlying multivariate process. Therefore,  there seem to be no relevant statistics for multivariate time series for which such a procedure will be valid.
 This is true even for the linear process class  (\ref{linearproc}), for which   the distribution of statistics like auto or  cross-correlations depends on the fourth-order moment structure of the process.
 %In other words, the aforementioned
 % extension of  the univariate multiplicative periodogram bootstrap to the multivariate context,   will be  valid in considerably fewer cases.
 In other words,    for multivariate time series, any frequency domain bootstrap scheme that ignores the weak dependence structure of the periodogram matrix across frequencies,  would be systematically failing to  properly capture the distribution of  a large  class of relevant  statistics.
  %SAY SOMETHING  HYBRID WILD and HYBRID FREQUENCY DOMAIN \\

%In this paper we propose 
To summarize,  
a frequency domain bootstrap method that is capable of handling
large classes of periodogram based statistics  
%the class of  integrated periodogram  statistics 
for general multivariate processes
 is not available so far. It is the purpose of the present paper to close this gap. 
Towards this end we introduce the concept of the \emph{multivariate  frequency domain hybrid  bootstrap} (MFHB). This procedure is composed of two ingredients. The first is the multivariate analogue of the previously mentioned multiplicative bootstrap approach proposed
%by Franke and H\"ardle (1992) and Dahlhaus and Janas (1996)
 for univariate processes. This ingredient is used to  imitate  features of the distribution of interest that  depend on the second-order structure of the process.  The second ingredient of our  MFHB  uses an adaptation
 of the convolved subsampling idea (cf.~Tewes et al.~(2019))  to the multivariate frequency domain context. It is used  to capture  those features of the same distribution that the first ingredient  is systematically missing out on, and which are crucially determined by the fourth-order structure of the process.
Notice that the MFHB procedure we propose  is structurally related to the hybrid procedure  that was proposed in Meyer et al.~(2020) for univariate time series. However, the MFHB for  multivariate time series  has to take a number of obstacles into account that are not present in the univariate situation and that have to be solved in a novel way.
%\footnote{MM: Is this too much of an overstatement of the differences between univariate and multivariate HPB?}. Firstly, the asymptotic distribution of periodogram ordinates at different frequencies no longer has a multiplicative structure in the multivariate context. Secondly, and even
In particular,    the limiting distribution of multivariate integrated periodogram statistics -- as well as of functions thereof -- is a multivariate complex normal distribution  where both covariance and relation matrix  depend  on the fourth-order moment structure of the  underlying  process. A  valid frequency domain bootstrap procedure therefore has to imitate correctly  not only the  covariance but also the  relation matrix of  the corresponding distribution.
%This is achieved in our procedure by appropriately designing the bootstrap scheme. 
Moreover, additional difficulties
arise  in the multivariate context from applying  a frequency domain bootstrap   to  the class of smooth functions of  integrated periodograms. Such an extension of the MFHB  turns out to be  important because
it  allows for applications to interesting classes of statistics, like  for instance,  to cross-correlations. In order to make  such an extension   possible,
 the second and the fourth-order characteristics of the covariance and  of the relation matrix  of the related  smooth functions  have  to be separated and appropriately imitated by the two ingredients of the MFHB procedure. As we will see, in contrast to integrated periodograms, this separation  can not be explicitly calculated and therefore the bootstrap scheme has to be  appropriately modified. We show that the proposed MFHB procedure
 simultaneously establishes correct covariance and relation matrix estimation for both aforementioned classes of statistics.
 % and that this is achieved under very general conditions on the class of processes and of statistics considered.
Moreover, the MFHB   will be proven to be valid for  integrated periodogram statistics  as well as for smooth functions thereof and for a  wide  class of stochastic processes which includes many known linear and nonlinear multivariate time series models.

 %In particular, it will be the first available bootstrap algorithm that is consistent for important statistics such as sample cross-covariances and cross-correlations under very general assumptions.\\
The remainder of this paper is organized as follows. Section 2 states the assumptions we impose on the multivariate process  and recalls some definitions and  limiting results for multivariate integrated periodogram statistics. 
%The important parameters of the corresponding limiting Gaussian distribution are  stated  and their dependence  on the fourth-order cumulants of the underlying vector process is  discussed.
Section 3 presents the MFHB procedure for integrated periodograms and for functions thereof, discusses some of its features and establishes its  asymptotic validity.  Section 4  presents  simulations that   investigate the finite sample performance of the MFHB and compares it with   that of  the moving block bootstrap.  The proofs of our main results are presented in the Appendix of this paper while the proofs of some technical results are deferred to the Supplementary Material.

\section{Preliminaries and basic results}

Consider  an $ \R^m $-valued, weakly stationary stochastic process $ ({\bf X}(t))_{t \in \Z} $ with mean zero and component processes  $ (X_{r}(t))_{t \in \Z} $, $ r=1,\ldots,m $. We denote the autocovariance matrix of the process by   $ \boldsymbol\Gamma(h)=E({\bf X}(t+h){\bf X}(t)^\top) \in \R^{m \times m} $, for $ h \in \Z $, with entries $ \gamma_{rs}(h)={\rm Cov}(X_{r}(t+h),X_{s}(t)) $, $ r,s \in \{1,\ldots,m\} $, which fulfill $ \gamma_{rs}(h)= \gamma_{sr}(-h) $.
Furthermore, the $k$-th order cumulant of $ ({\bf X}(t))_{t \in \Z} $ is denoted,  for any $  r_1, \ldots, r_k \in \{1, \ldots, m\} $ and $   t,  h_1, \ldots,  h_{k-1} \in \Z$,  by
\begin{equation} \label{eq.cum}
\textrm{cum}(X_{r_1}(t+h_1),X_{r_2}(t+h_2), \ldots, X_{r_{k-1}}(t+h_{k-1}), X_{r_{k}}(t)) \,.
\end{equation}

The following stationarity and weak dependence assumptions are  imposed on  the process under consideration.

\begin{assumption} \label{assu1}
$({\bf X}(t))_{t \in \Z}$  is eighth-order stationary, i.e. for all $ r_j \in \{1,2, \ldots, m\} $ all  joint cumulants of the process up to order eight
do not depend on the time point $ t $. We therefore write
$c_{r_1r_2\ldots r_k}(h_1, h_2, \ldots, h_{k-1})$ for \eqref{eq.cum}. Furthermore, it holds for all $ r_1,\ldots,r_8 \in \{1, \ldots, m\} $
\begin{enumerate}
\item[(i)]  \ $ \sum_{h\in \Z} (1+|h|) |\gamma_{r_1r_2}(h) | <\infty$,\\
%\item[]
\item[(ii)] \ $  \sum_{h_1,h_2,h_3 \in \Z}(1+|h_1| + |h_2| +|h_3|) |c_{r_1r_2r_3r_4}(h_1,h_2,h_3)| <\infty \,$,\\
%\item[]
\item[(iii)]\  $ \sum_{h_1, \ldots,h_7 \in \Z} |c_{r_1\ldots r_8}(h_1, h_2, \ldots, h_7) | <\infty \,.$
\end{enumerate}
\end{assumption}

The above assumption is satisfied for a large class of stochastic processes which includes, for instance, the multivariate
 linear processes  class (\ref{linearproc}) under certain moment assumptions on the innovations ${\bf e}_t$  and summability conditions on the matrices $ {\bf B}_j$; see Brillinger (1981).

Notice that since  $ \sum_{h\in \Z}|\gamma_{r_1 r_2}(h)| < \infty$, for all $ r_1,r_2 \in \{1,2,\ldots, m\}$,
the  process  $({\bf X}(t))_{t \in \Z}$ possesses a  spectral density matrix $ {\bf f}: [-\pi,\pi] \rightarrow \C^{m \times m} $ with entries
\bee
f_{rs}(\lambda) = \frac{1}{2\pi} \sum_{h \in \Z} \gamma_{rs}(h) \, e^{-ih\lambda}\,, \ \ \lambda\in [-\pi,\pi],
\eee
$r,s \in \{1,2, \ldots, m\}$, which is   Hermitian, i.e.,  $ f_{rs}(\lambda)= f_{sr}(-\lambda)= \overline{f_{sr}(\lambda)} $ and  $ \overline{\bf f}(\lambda) = {\bf f }(\lambda)$. Here, and throughout this work, $ \overline{\bf A} $ denotes the conjugate transpose of a complex-valued matrix $ {\bf A} $.
Furthermore,  ${\bf f}$ is bounded from above and is a continuous function of the frequency $\lambda$. In the following we additionally assume that
the eigenvalues of the spectral density matrix ${\bf f}$ are all bounded away from zero over  all frequencies.

\begin{assumption} \label{assu2}
A  constant $ \delta >0$ exists such that   the eigenvalues of the spectral density matrix, i.e.,  $ \sigma\big({\bf f}(\lambda)\big)$ satisfy
\[ \min \big( \sigma({\bf f}(\lambda)\big) \big)> \delta,\]
for all frequencies  $ \lambda \in (-\pi, \pi ]$.
\end{assumption}

Given an $m$-dimensional vector time series $ {\bf X}(1), {\bf X}(2), \ldots, {\bf X}(n)$  stemming from     $({\bf X}(t))_{t \in \Z}$,
 a common  moment estimator of the cross covariances   $ \gamma_{rs}(h)$, for $ -n< h < n$,  is given by
\bee
\widehat \gamma_{rs}(h)= \begin{cases}
\frac{1}{n} \sum_{t=1}^{n-h} X_{r}(t+h) X_{s}(t), & \textrm{ for } 0 \leq h \leq n-1\\
\frac{1}{n} \sum_{t=1}^{n-|h|} X_{r}(t) X_{s}(t+|h|), & \textrm{ for } -(n-1) \leq h \leq -1
\end{cases}\,.
\eee
In the following,  the  periodogram matrix $ {\bf I}: [-\pi,\pi] \rightarrow \C^{m \times m} $ with $ {\bf I}(\lambda)={\bf d}(\lambda) \, \overline{\bf d}(\lambda)$  is used as a basic statistic, where
\[ {\bf d}(\lambda)=(d_{1}(\lambda),\ldots,d_{m}(\lambda))^\top ,\]
 and the $m$-dimensional vector of finite  Fourier transforms $d_{r}$ is given by
\bee
d_{r}(\lambda)= \frac{1}{\sqrt{2\pi n}} \sum_{t=1}^{n} X_r(t) \, e^{-it\lambda}\,, \quad r=1,2, \ldots, m.
\eee
Notice that  $ I_{rs}(\lambda)=d_r(\lambda) \, d_s(-\lambda) $ while   $ I_{rr}(\lambda)I_{ss}(\lambda) =I_{rs}(\lambda)I_{sr}(\lambda) $.  Furthermore, it holds
\bee
I_{rs}(\lambda) = \frac{1}{2\pi} \sum_{h=-(n-1)}^{n-1} \widehat\gamma_{rs}(h) \, e^{-ih\lambda}\,.
\eee

Let $ \lambda_{j,n} =2\pi j/n$, $ j \in {\mathcal G}(n)$, be the Fourier frequencies based on a sample size $n$,  where
\be
\mathcal{G}(n) &:=& \{ j \in \Z: \; 1\leq |j| \leq [n/2] \}. \label{defGn}
\ee
Denote further   for $ p,q,r,s \in \{1,2, \ldots, m\}$, by
\bee
f_{pqrs}(\lambda,\mu,\eta)= \frac{1}{(2\pi)^3} \sum_{h_1,h_2,h_3\in \Z} c_{pqrs}(h_1,h_2,h_3) \, e^{-i(h_1\lambda+ h_2\mu + h_3\eta)}
\eee
 the fourth-order cumulant spectral density of $ ({\bf X}(t))_{t\in\Z}$.
The following lemma, which describes the covariance structure of the elements of the periodogram matrix at the Fourier frequencies, is very useful for our subsequent analysis.
  % Its generalizes to the  multivariate case   an univariate result
 %given in  Lemma XXX of Krogstad (XXX).
 %in the sequel a
%Regarding the covariance structure of the periodogram at the Fourier frequencies we have the following lemma:

\begin{lemma} \label{lemmacovper}
Let $ ({\bf X}(t))_{t \in \Z} $ have finite fourth moments and be fourth-order stationary satisfying the summability conditions (i) and (ii) of Assumption \ref{assu1}.
%, with the following summability assumptions on the cumulants:
%$$
%\sum_{h \in \Z} (1+|h|) \, |\gamma_{rs}(h)| < \infty$ and
%$$
%\sum_{h_1,h_2,h_3 \in \Z} (1+|h_1|+|h_2|+|h_3|) \, |c_{ r s v w}(h_1,h_2,h_3)| < \infty,$$
%for all $ r,s,v,w \in \{1,\ldots,m\} $.
Then it holds for all Fourier frequencies $ \lambda_{j,n},\lambda_{k,n} \in [0,\pi] $, and for all $ r,s,v,w \in \{1,\ldots,m\} $
\bee
{\rm Cov}(I_{rs}(\lambda_{j,n}),I_{vw}(\lambda_{k,n}))=S_1+S_2+S_3\,,
\eee
where
\begin{align*}
S_1 &= \frac{2\pi}{n} f_{rsvw}(\lambda_{j,n},-\lambda_{j,n},-\lambda_{k,n}) + \mathcal{O}\left(\frac{1}{n^2}\right)\,,\\
S_2 &= \begin{cases}
f_{rv}(\lambda_{j,n}) \cdot f_{sw}(-\lambda_{j,n}) + \mathcal{O}\left(\frac{1}{n}\right)\,, & \lambda_{j,n}=\lambda_{k,n}\\
\mathcal{O}\left(\frac{1}{n^2}\right)\,, & \lambda_{j,n} \neq \lambda_{k,n}
\end{cases}\,,\\
S_3 &= \begin{cases}
f_{rw}(\lambda_{j,n}) \cdot f_{sv}(\lambda_{j,n}) + \mathcal{O}\left(\frac{1}{n}\right)\,, & \lambda_{j,n}=\lambda_{k,n} \in \{0,\pi\}\\
\mathcal{O}\left(\frac{1}{n^2}\right)\,, & \textrm{else}
\end{cases}\,,
\end{align*}
and where all $ \mathcal{O}(\cdot) $ bounds are uniform over all Fourier frequencies.
%(\textbf{Note for us:} The $ -\lambda_{k,n} $ in the fourth-order term is correct, the minus vanishes later in the calculations! Also, this is for non-negative Fourier frequencies. For the negative ones, use $ I_{rs}(-\lambda)=I_{sr}(\lambda) $).
\end{lemma}

As it is seen from the above result, the covariance between the elements of the periodogram matrix at different frequencies vanishes by the order $1/n$ and depends on the fourth-order cumulant spectral density $f_{rsvw}$. We will see that this rate is not fast enough so that the covariance between periodogram ordinates -- and consequently the fourth-order structure of the process -- show up in the limiting distribution of integrated periodogram statistics. This  class of statistics,  which we will consider in the following, is defined as functions of the periodogram matrix. In particular, we require:

\begin{assumption} \label{assu3}
For some $ J \in \N $, $ \varphi_j: [-\pi,\pi] \rightarrow \C $, $ j=1,2,\ldots,J $ are square-integrable functions which are bounded in absolute value.
\end{assumption}

Now,  the vector of integrated  periodogram statistics we consider is  defined as
\begin{align}
{\bf M}_n= \left(M(\varphi_j,I_{r_js_j})=\int_{-\pi}^{\pi} \varphi_j(\lambda) I_{r_js_j}(\lambda) \, d\lambda\, , \,j =1,2, \ldots, J \right)^\top\,. \label{def_Mn}
\end{align}
Note that  ${\bf M_n} $   is an estimator of the  following  vector of  \emph{spectral means}
\begin{align}
{\bf M}= \left(M(\varphi_j,f_{r_js_j})=\int_{-\pi}^{\pi} \varphi_j(\lambda) f_{r_js_j}(\lambda) \, d\lambda\, , \,j =1,2, \ldots, J\right)^\top\,. \label{def_M}
\end{align}

Before proceeding with some limiting results regarding the behaviour of the estimators $ {\bf M}_n$, let us  look at some examples.

\begin{example}
The sample cross-covariance  $ \widehat\gamma_{rs}(h) $ at lag $ 0\leq  h < n $ is an integrated periodogram statistic. This is due to the fact that choosing $ \varphi(\lambda)=e^{ih\lambda} $ it follows from straightforward calculations that $ \widehat{\gamma}_{rs}(h)=M(\varphi,I_{rs}) $ as well as $ \gamma_{rs}(h)=M(\varphi,f_{rs}) $.
Notice that for $ -n < h <0$, $ \widehat\gamma_{rs}(h) = \widehat\gamma_{sr}(-h) $ is an estimator of $ \gamma_{rs}(h)$.
\end{example}
\begin{example} \label{examplesamplecorr}
The sample cross-correlation  $ \widehat\rho_{rs}(h)$ at lag $ 0\leq  h < n $ is a function  of integrated periodograms.
To elaborate,  let $ \varphi_1(\lambda)=e^{ih\lambda} $,  $ \varphi_2(\lambda)=\varphi_3(\lambda)=1$ and  consider the corresponding three-dimensional  vector of spectral means.
Then,
$$  \widehat\rho_{rs}(h)=M(\varphi_1,I_{rs})/\sqrt{M(\varphi_2,I_{rr})M(\varphi_3,I_{ss})}$$  is an estimator of
$  \rho_{rs}(h)=M(\varphi_1,f_{rs})/\sqrt{M(\varphi_2,f_{rr})M(\varphi_3,f_{ss})}$, the lag $h$ cross-correlation. Notice that $\rho_{rs}(h)$ and $ \widehat{\rho}_{rs}(h)$  are
functions
%$ g:\R^3\rightarrow \R$ with  $ g(x_1,x_2,x_3) = x_1/\sqrt{x_2 x_3}$  for  $ x_2 >0$, $x_3>0$,
of  the elements of   the  vectors
$  {\bf M}=(M(\varphi_1,f_{rs}), M(\varphi_2,f_{rr}), M(\varphi_3,f_{ss}))^\top$
 and $  {\bf M}_n=(M(\varphi_1,I_{rs}), M(\varphi_2,I_{rr}), M(\varphi_3,I_{ss}))^\top$, respectively. The limiting distribution of $ \widehat\rho_{rs}(h)$ will be derived in Example \ref{examplesamplecorrcontd}.
\end{example}

For practical calculations the integral in the expression for $ M(\varphi,I_{rs}) $ is commonly replaced by a Riemann sum using the Fourier frequencies.
%This is usually done using the Fourier frequencies based on sample size $ n $ which are given by $ \lambda_{j,n}=2\pi j/n $, for $ j \in \mathcal{G}(n) $, with
%\be
%\mathcal{G}(n) &:=& \{ j \in \Z: \; 1\leq |j| \leq [n/2] \}. \label{defGn}
%\ee
The corresponding approximation of $ M(\varphi,I_{rs}) $  is  given by
\bee
M_{\mathcal{G}(n)}(\varphi,I_{rs})= \frac{2\pi}{n} \sum_{l \in \mathcal{G}(n)} \varphi(\lambda_{l,n}) I_{rs}(\lambda_{l,n}) \,.
\eee
Before discussing the asymptotic properties of   ${\bf M}_n$,  we evaluate on some properties of the complex normal
distribution which are important for our subsequent discussion. An $ m $-dimensional complex random vector $ {\bf X}=(X_1, $ $ X_2,\ldots,X_m)^\top $ is called complex normal (or complex Gaussian) if and only if the $ 2m $-dimensional vector of real and imaginary parts
\bee
\big({\rm Re}({\bf X})^\top,{\rm Im}({\bf X})^\top \big)^\top := \big({\rm Re}(X_1),\ldots,{\rm Re}(X_m),{\rm Im}(X_1),\ldots,{\rm Im}(X_m) \big)^\top
\eee
has a $ 2m $-dimensional (real) normal distribution. The complex normal distribution is determined by three parameters: expectation $ \bm{\mu_X}=E({\bf X}) $, covariance matrix $ {\bf \Sigma_X} $, and relation matrix $ {\bf \Gamma_X} $, where
\bee
{\bf \Sigma_X}=E\big( [{\bf X}-\bm{\mu_X}] \, \overline{[{\bf X}-\bm{\mu_X}]} \big)\,, \quad {\bf \Gamma_X}=E\big( [{\bf X}-\bm{\mu_X}] \, [{\bf X}-\bm{\mu_X}]^\top \big)\,,
\eee
recalling that $ \overline {\bf A} $ denotes the conjugate transpose of any matrix or vector $ {\bf A} $. We therefore write
\bee
{\bf X} \sim \mathcal N_m^c(\bm{\mu_X},{\bf \Sigma_X},{\bf \Gamma_X})\,.
\eee
We are particularly interested in the case of centered vectors, i.e. $ \bm{\mu_X}={\bf 0} $. It is then easy to see that the covariance matrix of the real normal vector $ ({\rm Re}({\bf X})^\top,{\rm Im}({\bf X})^\top)^\top $ can be directly deduced from the parameters of the complex normal vector $ {\bf X} $ via
\be
{\bf X} \sim \mathcal N_m^c({\bf 0},{\bf \Sigma_X},{\bf \Gamma_X}) \quad \Leftrightarrow \quad \begin{pmatrix} {\rm Re}({\bf X})\\ {\rm Im}({\bf X})
\end{pmatrix} \sim \mathcal N_{2m}({\bf 0},{\bf G_X})\,, \label{equiv_complexreal}
\ee
where
\begin{align*}
{\bf G_X} & = E\left[\begin{pmatrix}{\rm Re}({\bf X})\\ {\rm Im}({\bf X})\end{pmatrix}\begin{pmatrix}{\rm Re}({\bf X})\\ {\rm Im}({\bf X})\end{pmatrix}^\top \right] \\
&= \left(\begin{array}{cc}  \frac{1}{2}({\rm Re}({\bf \Sigma_X} )+{\rm Re}({\bf \Gamma_X}))&    \frac{1}{2}(-{\rm Im}({\bf \Sigma_X}) +{\rm Im}({\bf \Gamma_X}))\\
& \\
\frac{1}{2}({\rm Im}({\bf \Sigma_X}) +{\rm Im}({\bf \Gamma_X})) &    \frac{1}{2}({\rm Re}({\bf \Sigma_X}) -{\rm Re}({\bf \Gamma_X} ))
\end{array} \right)\,.
\end{align*}
The above expression shows that $ {\bf G_X} $ is fully determined by $ {\bf \Sigma_X}$ and $ {\bf \Gamma_X}$ and vice versa. In particular, denote  by ${\bf G}_{{\bf X};ij}$, $ i,j\in \{1,2\}$, the four matrices  appearing in the $i$-th row and $j$-th column of the above block matrix $ {\bf G_X}$. Then   $ {\bf \Sigma_X}= ({\bf G}_{{\bf X};11}+{\bf G}_{{\bf X};22}) + i\, ({\bf G}_{{\bf X};21} -{\bf G}_{{\bf X};12})$ while
$ {\bf \Gamma_X} =  ({\bf G}_{{\bf X};11}-{\bf G}_{{\bf X};22}) + i\, ({\bf G}_{{\bf X};21} +{\bf G}_{{\bf X};12})$.\\
The equivalence from \eqref{equiv_complexreal} also carries over to weak convergence: For a sequence of complex random vectors $ ({\bf X}_n)_{n \in \N} $ we have $ {\bf X}_n \stackrel{d}{\rightarrow} \mathcal N_m^c({\bf 0},{\bf \Sigma_X},{\bf \Gamma_X}) $ if and only if $ ({\rm Re}({\bf X}_n)^\top,{\rm Im}({\bf X}_n)^\top)^\top \stackrel{d}{\rightarrow} \mathcal N_{2m}({\bf 0},{\bf G_X}) $.\\
For $ \bm{\mu_X}={\bf 0} $ there are two particularly important special cases where the distribution is completely determined by the matrix $ {\bf \Sigma_X} $, the real normal case and the circularly symmetric case. The centered random vector $ {\bf X} $ is almost surely real-valued if and only  if $ {\bf \Sigma_X}={\bf \Gamma_X} $, as a simple calculation shows. In this case one may switch to the usual notation for real-valued random vectors:
\bee
{\bf X} \sim \mathcal N_m^c({\bf 0},{\bf \Sigma_X},{\bf \Sigma_X}) \quad \Leftrightarrow \quad {\bf X} \sim \mathcal N_{m}({\bf 0},{\bf \Sigma_X})\,.
\eee
Another important special case is the circularly symmetric case. $ {\bf X} $ is called circularly symmetric if for all $ \varphi \in (-\pi,\pi] $ the distribution of $ e^{i\varphi} {\bf X} $ equals the distribution of $ {\bf X} $. This is the case if and only if $ \bm{\mu_X}={\bf 0} $ and $ {\bf \Gamma_X}={\bf 0} $. In this case the joint distribution of real and imaginary parts takes a very specific  form, as can be seen from \eqref{equiv_complexreal}:
\bee
{\bf X} \sim \mathcal N_m^c({\bf 0},{\bf \Sigma_X},{\bf 0})
\Leftrightarrow  \begin{pmatrix}
{\rm Re}({\bf X})\\
{\rm Im}({\bf X})
\end{pmatrix} \sim \mathcal N_{2m} \left( \begin{pmatrix}
{\bf 0}\\
{\bf 0}
\end{pmatrix} \,, \frac{1}{2}\begin{pmatrix}
{\rm Re}({\bf \Sigma_X}) & -{\rm Im}({\bf \Sigma_X})\\
{\rm Im}({\bf \Sigma_X}) & {\rm Re}({\bf \Sigma_X})
\end{pmatrix} \right)\,.
\eee
In particular, $ {\rm Re}({\bf X}) $ and $ {\rm Im}({\bf X}) $ then are identically distributed.

\vspace*{0.2cm}

Under  Assumption \ref{assu1} and some  additional weak dependence conditions on the process $ ({\bf X}(t))_{t\in\Z}$ it is known that, for any $ r,s \in \{1,\ldots,m\} $, $ M(\varphi,I_{rs}) $ is a consistent estimator for $ M(\varphi,f_{rs}) $ and that the following central limit theorem holds true for $ {\bf V}_n:=\sqrt{n}({\bf M}_n-{\bf M})$:
\begin{align}
{\bf V}_n=\sqrt{n} \Big( M(\varphi_j,I_{r_j s_j})-M(\varphi_j,f_{r_j s_j})\, , \, j=1,\ldots,J \Big)^\top \stackrel{d}{\longrightarrow} {\bf V}\,,  \label{CLT}
\end{align}
where $ {\bf V} $ is a $ J $-dimensional complex normal random vector with mean zero, covariance matrix $ {\bf \Sigma} $, and relation matrix $ {\bf \Gamma} $.
That is,
%We write
\bee
{\bf V}=(V_{r_j s_j}(\varphi_j), \, j=1,\ldots,J)^\top \sim \mathcal N_J^c({\bf 0},{\bf \Sigma},{\bf \Gamma}).
\eee
 $ {\bf V} $ fulfils
\be
\overline{V_{r_j s_j}(\varphi_j(\cdot))} = V_{r_j s_j}(\overline{\varphi_j(- \,\cdot)})\,, \label{limit_prop}
\ee
and covariance and relation parameters are given as follows. The covariance matrix decomposes into $ {\bf \Sigma}={\bf \Sigma}_1+{\bf \Sigma}_2 $ where the $(j,k)$-th element of the matrix ${\bf \Sigma}$  is given by
\be
\Sigma_{j k}=\textrm{Cov}(V_{r_j s_j}(\varphi_j),V_{r_k s_k}(\varphi_k)) = \Sigma_{1;j k} + \Sigma_{2;j k}  \label{covrelformula}
\ee
with
\begin{align}
\Sigma_{1;j k} &= 2\pi  \int_{-\pi}^{\pi} \varphi_j(\lambda) \overline{\varphi_k(\lambda)} f_{r_j r_k}(\lambda) f_{s_j s_k}(-\lambda) \, d\lambda \label{limit_var1} \\
& \quad \quad + 2\pi \int_{-\pi}^{\pi} \varphi_j(\lambda) \overline{\varphi_k(-\lambda)} f_{r_j s_k}(\lambda) f_{s_j r_k}(-\lambda) \, d\lambda\, \,, \nonumber
\end{align}
and
\begin{align}
\Sigma_{2; jk} &= 2\pi \, \int_{-\pi}^{\pi} \int_{-\pi}^{\pi} \varphi_j(\lambda_1) \overline{\varphi_k(-\lambda_2)} f_{r_j s_j r_k s_k}(\lambda_1,-\lambda_1,\lambda_2) \, d\lambda_1 \, d\lambda_2 \,. \label{limit_var2}
\end{align}
%Here,
%\bee
%f_{pqrs}(\lambda,\mu,\eta)= \frac{1}{(2\pi)^3} \sum_{h_1,h_2,h_3\in \Z} c_{pqrs}(h_1,h_2,h_3) \, e^{-i(h_1\lambda+ h_2\mu + h_3\eta)}
%\eee
%is the fourth-order cumulant spectral density of $ ({\bf X}(t))_{t\in\Z}$.
The relation matrix decomposes into $ {\bf \Gamma}={\bf \Gamma}_1+{\bf \Gamma}_2 $ where the $(j,k)$-th element is given by
\be
\Gamma_{j k}=\textrm{Cov}(V_{r_j s_j}(\varphi_j),\overline{V_{r_k s_k}(\varphi_k)}) = \Gamma_{1;j k} + \Gamma_{2;j k}  \label{covrelformula_rel}
\ee
with
\begin{align}
\Gamma_{1;j k} &= 2\pi  \int_{-\pi}^{\pi} \varphi_j(\lambda) \varphi_k(-\lambda) f_{r_j r_k}(\lambda) f_{s_j s_k}(-\lambda) \, d\lambda \label{limit_rel1} \\
& \quad \quad + 2\pi \int_{-\pi}^{\pi} \varphi_j(\lambda) \varphi_k(\lambda) f_{r_j s_k}(\lambda) f_{s_j r_k}(-\lambda) \, d\lambda\, \,, \nonumber
\end{align}
and
\begin{align}
\Gamma_{2; jk} &= 2\pi \, \int_{-\pi}^{\pi} \int_{-\pi}^{\pi} \varphi_j(\lambda_1) \varphi_k(\lambda_2) f_{r_j s_j r_k s_k}(\lambda_1,-\lambda_1,\lambda_2) \, d\lambda_1 \, d\lambda_2 \,. \label{limit_rel2}
\end{align}
Observe that due to \eqref{limit_prop} it holds $ \Gamma_{j k} = \textrm{Cov}(V_{r_j s_j}(\varphi_j(\cdot)),V_{r_k s_k}(\overline{\varphi_k(- \,\cdot)})) $. Hence, $ {\bf \Gamma} $ can be obtained from $ {\bf \Sigma} $ by replacing $ \varphi_k(\cdot) $ with $ \overline{\varphi_k(- \,\cdot)} $, and we stated the explicit form merely for convenience reasons. We refer to Rosenblatt (1963),  Brillinger (1981), Dahlhaus (1985) and  Taniguchi and  Kakizawa (2000).
%Here, the cumulants are defined via
%\bee
%c_{pqrs}(h_1,h_2,h_3)= {\rm cum}(X_p(h_1),X_q(h_2),X_r(h_3),X_s(0))\,.
%\eee
As it is seen from the above expressions the terms ${\bf \Sigma}_2$ and ${\bf \Gamma}_2$ appearing in the covariance and relation matrices of the limiting  complex Gaussian distribution
depend on the entire fourth-order moment structure of the process $({\bf X}(t))_{t\in\Z}$, and this dependence is due to the covariance  of  periodogram ordinates across frequencies; see Lemma~\ref{lemmacovper}.

As argued in \eqref{equiv_complexreal}, the weak convergence $ {\bf V}_n \stackrel{d}{\rightarrow} {\bf V}$ stated in (\ref{CLT}) is equivalent to the statement that the $2J$-dimensional real random vector
$ \big({\rm Re}({\bf V}_n)^\top , ({\rm Im}({\bf V}_n)^\top\big)^\top$ converges
weakly to a $ 2J$-dimensional real normal distribution with mean zero and covariance matrix
\begin{align}
 {\bf G} & = E\big[\left(\begin{array}{c}{\rm Re}({\bf V})\\ {\rm Im}({\bf V})\end{array}\right)\big({\rm Re}({\bf V})^\top,  {\rm Im}({\bf V})^\top \big)\big] \nonumber\\
 &= \left(\begin{array}{cc}  \frac{1}{2}({\rm Re}({\bf \Sigma} )+{\rm Re}({\bf \Gamma}))&    \frac{1}{2}(-{\rm Im}({\bf \Sigma}) +{\rm Im}({\bf \Gamma}))\\
 & \\
   \frac{1}{2}({\rm Im}({\bf \Sigma}) +{\rm Im}({\bf \Gamma})) &    \frac{1}{2}({\rm Re}({\bf \Sigma}) -{\rm Re}({\bf \Gamma} ))
 \end{array} \right)\,. \label{defG}
 \end{align}

Notice that the limiting distribution $ {\bf V}$ is not necessarily real-valued even  if the functions $ \varphi_j, j=1,2, \ldots, J$ are real-valued.   However, if  the functions $\varphi_j$ satisfy $ \varphi_j(-\lambda)= \overline{\varphi_j(\lambda)}$ for $ j=1,2, \ldots, J$,  then $ {\bf \Sigma}={\bf \Gamma}$ and  $ {\bf V}$  is real valued, that is $  {\bf V}\sim {\mathcal N}_J({\bf 0}, {\bf \Sigma})$.

\begin{example}[Ex.~\ref{examplesamplecorr} continued] \label{examplesamplecorrcontd}
With the central limit theorem given for $ {\bf V}_n $ above, we can now state the limiting distribution of the sample cross-correlation. For the particular vectors $ {\bf M}_n $ and $ {\bf M} $ from Example \ref{examplesamplecorr} it can easily be seen that $ {\bf V}_n $ is real-valued and converges to a real Gaussian random vector $ {\bf V} $. Applying the delta method to this CLT it follows after straightforward but tedious computations that
\bee
\sqrt{n} \left( \widehat \rho_{rs}(h) - \rho_{rs}(h) \right) \stackrel{d}{\longrightarrow} \mathcal N(0,\tau^2)\,,
\eee
where
\begin{align*}
\tau^2 = \sum_{j \in \Z} &\left\{ \rho_{rr}(j)\rho_{ss}(j) + \rho_{rs}(j+h) \rho_{sr}(j-h) + \frac{c_{rsrs}(j,j-h,h)}{\gamma_{rr}(0) \gamma_{ss}(0)}  \right. \\
&+ \frac{1}{2}\rho_{rs}(h)^2 \big[ \rho_{rr}(j)^2 + \rho_{ss}(j)^2 + 2\rho_{rs}(j)^2 \big] \\
&- 2\rho_{rs}(h) \big[ \rho_{rr}(j)\rho_{sr}(j-h) + \rho_{rs}(j)\rho_{ss}(j-h) \big] \\
&+ \frac{1}{4} \rho_{rs}(h)^2 \left[ \frac{c_{rrrr}(j,j,0)}{\gamma_{rr}(0)^2} + \frac{c_{ssss}(j,j,0)}{\gamma_{ss}(0)^2} + \frac{2 \, c_{rrss}(j,j,0)}{\gamma_{rr}(0) \gamma_{ss}(0)} \right] \\
&- \left.\frac{\rho_{rs}(h)}{\sqrt{\gamma_{rr}(0) \gamma_{ss}(0)}} \left[ \frac{c_{rsrr}(j,j-h,0)}{\gamma_{rr}(0)} + \frac{c_{rsss}(j,j-h,0)}{\gamma_{ss}(0)} \right] \right\} \,.
\end{align*}
\end{example}

The  above dependence of the  variance of the limiting Gaussian distribution on the fourth-order structure of the process does not simplify even in the case of linear processes as this is the case for univariate linear processes. The following simple example  illustrates this fact.
%
%As the considerations in Section 1 show, for multivariate stochastic processes, the limiting distribution of periodogram-based statistics
%is much more complicated since it  depends on the entire second and  fourth-order moment structure of the underlying process. This is in contrast to the univariate case where for particular classes of statistics and processes, like for instance for autocorrelations for linear processes, the fourth-order structure of the process does not show up in the limiting distribution of the corresponding sample estimates. The following example illustrates this fact.

\begin{example}[Example~\ref{examplesamplecorrcontd} continued] \label{examplefourthorder}
%Sample cross-correlations and autocorrelations are special cases of the statistics under investigation in this paper, as was argued in Examples \ref{examplesamplecorr} and \ref{examplesamplecorrcontd}. Note that the limiting variance $ \tau^2 $ in Example \ref{examplesamplecorrcontd} depends on the fourth-order cumulants of the underlying process. This is also true for the special case of \emph{univariate} time series ($ m=r=s=1 $). However, if one considers \emph{linear univariate} time series, i.e. moving averages driven by an \emph{i.i.d.}~white noise process, $ \tau^2 $ simplifies to a form where all fourth-order terms cancel out and only the (second-order) autocorrelation structure shows up in the limiting variance, cf. \citet{BrockwellDavis91}, Theorem 7.2.1.\\
%This nice feature of univariate linear processes does no longer hold true in the multivariate context. As an example,
Consider the following simple bivariate linear process, which is a vector moving average process of order $ 1 $:
\be
{\bf X}(t)=\begin{pmatrix}
X_1(t)\\
X_2(t)
\end{pmatrix} = \begin{pmatrix}
\varepsilon_1(t)\\
\varepsilon_2(t)
\end{pmatrix} + \begin{pmatrix}
1 & 1\\
1 & -1
\end{pmatrix} \, \begin{pmatrix}
\varepsilon_1(t-1)\\
\varepsilon_2(t-1)
\end{pmatrix}\,, \label{exampleprocess}
\ee
where $ (\varepsilon_1(t))_{t \in \Z} $ and $ (\varepsilon_2(t))_{t \in \Z} $ are independent univariate i.i.d.~white noise processes, with $ E(\varepsilon_j(t))=0 $, $ {\rm E}(\varepsilon_j(t)^2)=1 $, and kurtosis ${\rm  E}(\varepsilon_j(t)^4)=: \eta_j $, for $ j=1,2 $. We take a look at the cross-correlation at lag $ h=0 $, and derive the limiting variance of $ \sqrt{n} ( \widehat \rho_{12}(0) - \rho_{12}(0) ) $. For the process \eqref{exampleprocess} it holds $ \rho_{12}(0)=0 $, and the expression $ \tau^2 $ from Example \ref{examplesamplecorrcontd} simplifies considerably to
\begin{align*}
\tau^2 &= \rho_{11}(0)\rho_{22}(0) + 2\rho_{11}(1)\rho_{22}(1) + 2\rho_{12}(1)\rho_{21}(1) + \sum_{j \in \Z} \frac{c_{1212}(j,j,0)}{\gamma_{11}(0) \gamma_{22}(0)} \\
&= 1+ \frac{\eta_1-3}{9} + \frac{\eta_2-3}{9}\,.
\end{align*}
As it is seen -- and   in contrast to what happens for univariate linear processes -- in the multivariate context the limiting variance depends on the fourth-order structure of the underlying white noise which can in general not be expressed in terms of second-order quantities of the process $ ({\bf X}(t)) $.
\end{example}

The considerations in this section made it clear that   for a frequency domain bootstrap to be successful  in the multivariate context, it has to appropriately imitate the second and  the fourth order structure of  the underlying stochastic process.  This is what  the MFHB procedure achieves.
%presented in the next section.
%To achieve this goal we propose a hybrid type procedure which essentially uses a  resampling from the appropriate Wishart distribution to imitate the second-order structure  and a modified convolved subsampling procedure to deliver the missing fourth-order terms which show up in  the weak covariance structure of the periodogram matrix.

\section{The multivariate frequency domain hybrid bootstrap (MFHB)}
We discuss the frequency domain procedure  proposed in this paper in two parts. First  we motivate and describe the MFHB procedure for integrated periodograms.
Then  we present a modification  of the  MFHB so that in can be successfully applied to functions of integrated periodograms.

\subsection{The MFHB  for integrated periodograms} \label{subsec_procedure}
 The  MFHB procedure   for  integrated periodograms generates  two sets of bootstrap pseudo random variables  which will be denoted by the superscripts $ \ast $ and $ + $, respectively.  The procedure is divided
 into three main steps which  are denoted  by Step I, Step II and Step III. In Step I   independent pseudo periodogram matrices are generated  which are  denoted by $ {\bf I}^\ast$. This is  done  using the asymptotic complex Gaussian distribution  of the  vector of finite Fourier transforms and their  asymptotic independence  across frequencies. In Step II the idea of convolved bootstrap of subsamples, cf.~Tewes et al.~(2019), is adopted  to develop an algorithm which  generates a second  set of  pseudo periodogram matrices,  denoted by $ {\bf I}^{+}$, which are independent of ${\bf I}^\ast$.
 The pseudo periodogram matrices  $ {\bf I}^+$ correctly  imitate  the weak dependence of the  ordinary  periodogram matrices $ {\bf I}$ across  frequencies within subsamples.
 Step III merges the integrated periodogram statistics based on the two bootstrapped periodograms $ {\bf I}^\ast$ and $ {\bf I}^+$  in an appropriate way. The  merging ensures that replicates of the integrated periodograms based on $ {\bf I}^\ast$   imitate all features of the corresponding distribution  up to those depending on the fourth-order structure of the process. The fourth-order features of this distribution  are  contributed  by the corresponding statistics  based on the pseudo periodograms $ {\bf I}^+$.  Notice that the MFHB bootstrap approximations are designed
 in such a way  that the covariance {\it and}  the relation matrix
of the limiting complex Gaussian distribution of the integrated periodograms is consistently estimated, see Remark \ref{rem32}.
Furthermore,   the bootstrap random  matrices   $ {\bf I}^\ast $ and $ {\bf I}^+ $ are defined on the same probability space, with probability measure $ P^\ast $, and are independent from each other. Consequently,  all  bootstrap expectations, variances and covariances are denoted in the following   by $ {\rm E}^\ast $, $ {\rm Var}^\ast $, and $ {\rm Cov}^\ast $, respectively.

The following algorithm   implements  the previously discussed ideas.

\begin{enumerate}
\item[] {\bf Step I.1} \  Let $ \widehat{\bf f}$ be an estimator of the spectral density matrix $ {\bf f}$ and denote by $\widehat{f}_{rs}(\lambda) $ the $ (r,s)$-th element of $ \widehat{\bf f}(\lambda)$.
\item[] {\bf Step I.2} \ Generate, independently for $ j=1,2, \ldots,N:=[n/2]$, pseudo Fourier transforms
\[ {\bf d}^\ast(\lambda_{j,n})=(d^\ast_{1}(\lambda_{j,n}), d^\ast_{2}(\lambda_{j,n}), \ldots, d_{m}^\ast(\lambda_{j,n}))^\top   \sim {\mathcal N}_m^c({\bf 0}, \widehat{\bf f}(\lambda_{j,n}), {\bf 0})\]
and  calculate the pseudo periodogram matrices
\bee {\bf  I}^\ast(\lambda_{j,n}) = \begin{cases}
 {\bf d}^\ast(\lambda_{j,n})\overline{\bf d}^\ast(\lambda_{j,n}) &  \mbox{for} \  j=1,2, \ldots, N\,,\\
 {\bf I}^\ast(-\lambda_{j,n})^\top, & \mbox{for} \ j=-1, -2, \ldots, -N \end{cases}
\eee
with entries  denoted by $ I^\ast_{rs}(\lambda_{j,n}) $, $ r,s\in\{1,2, \ldots, m\}$.
\item[] {\bf Step I.3} \ Let
\[ {\bf V}^\ast_n = \sqrt{n} \Big( M_{{\mathcal G}(n)}(\varphi_j,I^\ast_{r_j,s_j} )-M_{{\mathcal G}(n)}(\varphi_j,\widehat{f}_{r_j,s_j}) \, , \, j=1, 2, \ldots, J\Big)^\top,\]
be the analogue of $ {\bf V}_n$ based on $ {\bf I}^\ast$.
% and  let $ {\bf \Sigma}^\ast_n=\textrm{Var}^\ast({\bf V}^\ast_n)$.
\item[] {\bf Step II.1} \ Select a positive integer $ b < n$  and consider the set of all periodogram matrices  based on subsamples of length $b$. Denote by
 $ {\bf I}_t(\lambda_{j,b})$  the periodogram matrix  of the subsample  $ {\bf X}(t), {\bf X}(t+1), \ldots, {\bf X}(t+b-1)$ at frequency  $\lambda_{j,b} = 2\pi j/b$,  $\lambda_{j,b}  \in {\mathcal G}(b)$.
Let $ \widetilde{\bf f}(\lambda_{j,b})= (n-b+1)^{-1}\sum_{t=1}^{n-b+1} {\bf I}_t(\lambda_{j,b})$ be the average of the periodogram matrices of the  subsamples at frequency $ \lambda_{j,b} $.
\item[] {\bf Step II.2} \  Define the  set of frequency domain  residual matrices
$$ {\bf U}_{t}(\lambda_{j,b})  = \widetilde{\bf f}^{-1/2}(\lambda_{j,b}){\bf I}_{t}(\lambda_{j,b})  \widetilde{\bf f}^{-1/2}(\lambda_{j,b}),$$
where  $j=1,2, \ldots, [b/2]$,  $t =1,2, \ldots, n-b+1$ and $ {\bf A}^{-1/2}$ denotes the square root of the  inverse  matrix ${\bf A}^{-1}$, i.e.,  ${\bf A}^{-1/2}{\bf A}^{-1/2}={\bf A}^{-1}$.
\item[] {\bf Step II.3} \ Let $ k=[n/b]$ and generate i.i.d.~bootstrap random variables $ i_1, i_2, \ldots, i_k$ with a discrete uniform distribution on the set
$ \{1,2, \ldots, n-b+1\}$. Let
\[ {\bf I}^+(\lambda_{j,b}) = \frac{1}{k}\sum_{\ell=1}^k {\bf I}^+_{i_\ell}(\lambda_{j,b}),\]
 where  $ {\bf I}^+_{i_\ell}(\lambda_{j,b}) =  \widehat{\bf f}^{1/2}(\lambda_{j,b}){\bf U}_{i_\ell}(\lambda_{j,b})  \widehat{\bf f}^{1/2}(\lambda_{j,b}) $.
 We write  $ {\bf I}^+_{\ell}(\lambda_{j,b}) $ for $ {\bf I}^+_{i_\ell}(\lambda_{j,b})$.
The entries of the
 pseudo periodogram matrices  $ {\bf I}^+(\lambda_{j,b}) $ are denoted $ I^+_{rs}(\lambda_{j,b}) $.
 \item[] {\bf Step II.4} \ Let
 \[  {\bf V}^+_n = \sqrt{kb} \Big(M_{{\mathcal G}(b)}(\varphi_j,I^+_{r_j,s_j} )-M_{{\mathcal G}(b)}(\varphi_j,\widehat{f}_{r_j,s_j}) \, , \, j=1, 2, \ldots, J\Big)^\top\]
 be the analogue of $ {\bf V}_n$ based on $ {\bf I}^+$.
% where $ \widehat{\bf f}(\lambda_{j,b})= (\widehat f_{rs}(\lambda_{j,b}))_{r,s=1,\ldots,m} $.
 %:= (n-b+1)^{-1}\sum_{t=1}^{n-b+1} {\bf I}_{t}(\lambda_{j,b})$
 %and $ {\bf I}_{t}(\lambda_{j,b}) $ is the periodogram matrix of the subsample  $ {\bf X}(t), {\bf X}(t+1), \ldots, {\bf X}(t+b-1)$.
 %Let
 %${\bf \Sigma}_n^+ = \textrm{Var}^\ast({\bf V}^+_n)$.
 \item[] {\bf Step III.1} \  Let
 \[ {\bf G}_n^\ast = E^\ast\big[\left(\begin{array}{c}{\rm Re}({\bf V}^*_n)\\ {\rm Im}({\bf V}^*_n)\end{array}\right)\big({\rm Re}({\bf V}^*_n)^\top,  {\rm Im}({\bf V}^*_n)^\top \big)\big], \]
 \[ {\bf G}_n^+= E^\ast\big[\left(\begin{array}{c}{\rm Re}({\bf V}^+_n)\\ {\rm Im}({\bf V}^+_n)\end{array}\right)\big({\rm Re}({\bf V}^+_n)^\top,  {\rm Im}({\bf V}^+_n)^\top \big)\big].\]
 Furthermore, denote by $ {\bf \Sigma}_{1,n}^+ $  the $ J\times J$ matrix with $(j,k)$-th element given by
  \begin{align*}
    \sigma^+_{jk} & = \frac{4\pi^2}{b} \sum_{l=-[b/2]}^{[b/2]}\varphi_j(\lambda_{l,b}) \overline{\varphi_k(\lambda_{l,b})} S_{r_j s_js_kr_k}(\lambda_{l,b})\\
    % \widetilde{f}_{r_jr_k}(\lambda_{l,b}) \widetilde{f}_{s_js_k}(-\lambda_{l,b})\\
 & \ \ \ \ + \frac{4\pi^2}{b} \sum_{l=-[b/2]}^{[b/2]}\varphi_j(\lambda_{l,b}) \overline{\varphi_k(-\lambda_{l,b})} S_{r_js_jr_ks_k}(\lambda_{l,b}).
  %\sigma^+_{jk} & = 2\pi \Big( \frac{2\pi}{kb} \sum_{l=-[b/2]}^{[b/2]}\varphi_j(\lambda_{l,b}) \overline{\varphi_k(\lambda_{l,b})} \widetilde{f}_{r_jr_k}(\lambda_{l,b}) \widetilde{f}_{s_js_k}(-\lambda_{l,b})\\
 %& \ \ \ \ \ \  \ \ \ \ \ + \frac{2\pi}{kb} \sum_{l=-[b/2]}^{[b/2]}\varphi_j(\lambda_{l,b}) \overline{\varphi_k(-\lambda_{l,b})} \widetilde{f}_{r_js_k}(\lambda_{l,b}) \widetilde{f}_{s_jr_k}(-\lambda_{l,b})\Big).
\end{align*}
and $ {\bf \Gamma}_{1,n}^+ $  the $ J\times J$ matrix with $(j,k)$-th element given by
  \begin{align*}
  c^+_{jk} & = \frac{4\pi^2}{b}  \sum_{l=-[b/2]}^{[b/2]}\varphi_j(\lambda_{l,b}) \varphi_k(-\lambda_{l,b}) S_{r_js_j s_k r_k}(\lambda_{l,b})\\
 & \ \ \ \  + \frac{4\pi^2}{b} \sum_{l=-[b/2]}^{[b/2]}\varphi_j(\lambda_{l,b}) \varphi_k(\lambda_{l,b}) S_{r_js_jr_ks_k}(\lambda_{l,b}),
\end{align*}
where
%\[ S_{rsuw}(\lambda) = \frac{1}{n-b+1}\sum_{t=1}^{n-b+1}\big( I_{t,rs}(\lambda)-\widetilde{f}_{rs}(\lambda)\big) \big( I_{t,uw}(\lambda) - \widetilde{f}_{uw}(\lambda) \big),\]
\[ S_{rsuw}(\lambda) = \frac{1}{n-b+1}\sum_{t=1}^{n-b+1}\big( \widetilde{I}_{t,rs}(\lambda)-\widehat{f}_{rs}(\lambda)\big) \big( \widetilde{I}_{t,uw}(\lambda) - \widehat{f}_{uw}(\lambda) \big),\]
and $ \widetilde{I}_{t,rs}(\lambda)$ is the $(r,s)$-th element of the  matrix
$$ \widetilde{\bf I}_{t}(\lambda) =  \widehat{\bf f}^{1/2}(\lambda){\bf U}_{t}(\lambda)  \widehat{\bf f}^{1/2}(\lambda). $$
%  calculated using  the subsample $ {\bf X}(t), {\bf X}(t+1), \ldots, {\bf X}(t+b-1)$.
Finally, define the matrix
\[ {\bf C}^+_n = \left(\begin{array}{cc}  \frac{1}{2}({\rm Re}({\bf \Sigma}^+_{1,n} )+{\rm Re}({\bf \Gamma}^+_{1,n}))  &    \frac{1}{2}(-{\rm Im}({\bf \Sigma}^+_{1,n}) +{\rm Im}({\bf \Gamma}^+_{1,n}))\\
 & \\
   \frac{1}{2}({\rm Im}({\bf \Sigma}^+_{1,n}) +{\rm Im}({\bf \Gamma}^+_{1,n}))&    \frac{1}{2}({\rm Re}({\bf \Sigma}^+_{1,n}) -{\rm Re}({\bf \Gamma}^+_{1,n}))
 \end{array} \right).\]
 \item[] {\bf Step III.2}
Calculate
\begin{equation*}
 {\bf G}_n^\circ = {\bf G}_n^\ast + \big({\bf G}_n^+ - {\bf C}_n^+\big)
 \end{equation*}
and
\begin{equation}
\left(\begin{array}{c}{\rm Re}({\bf V}^\circ_n)\\ {\rm Im}({\bf V}^\circ_n)\end{array}\right) =  \big({\bf G}_{n}^\circ \big)^{1/2}\big({\bf G}^\ast_n \big)^{-1/2}
\left(\begin{array}{c}{\rm Re}({\bf V}^*_n)\\ {\rm Im}({\bf V}^*_n)\end{array}\right)\,. \label{eq.Co}
\end{equation}
\item[] {\bf Step III.3} Approximate the distribution of $ {\bf V}_n$ by that of
\[ {\bf V}_n^\circ = {\rm Re}({\bf V}^\circ_n)+ i \,{\rm Im}({\bf V}^\circ_n)\,.\]
\end{enumerate}

\vspace*{0.3cm}

The following series of  remarks clarifies several aspects of  the above bootstrap procedure.

\begin{remark} \label{rem_gen_cn}
In Step I.2, the problem of generating a complex normal random vector $ {\bf d}^\ast(\lambda_{j,n}) $ can in practice be reduced to that of generating a real normal random vector (which is usually a pre-implemented routine). Since $ {\mathcal N}_m^c({\bf 0}, \widehat{\bf f}(\lambda_{j,n}), {\bf 0}) $ is a circularly symmetric complex normal distribution, one may generate
\bee
 \begin{pmatrix}
\textrm{Re}({\bf d}^\ast(\lambda_{j,n}))\\
\textrm{Im}({\bf d}^\ast(\lambda_{j,n}))
\end{pmatrix} \sim {\mathcal N}_{2m} \left( \begin{pmatrix}
{\bf 0}\\
{\bf 0}
\end{pmatrix} \,, \frac{1}{2}\begin{pmatrix}
\textrm{Re}(\widehat{\bf f}(\lambda_{j,n})) & -\textrm{Im}(\widehat{\bf f}(\lambda_{j,n}))\\
\textrm{Im}(\widehat{\bf f}(\lambda_{j,n})) & \textrm{Re}(\widehat{\bf f}(\lambda_{j,n}))
\end{pmatrix} \right)\,,
\eee
and then set $ {\bf d}^\ast(\lambda_{j,n})=\textrm{Re}({\bf d}^\ast(\lambda_{j,n})) + i \cdot \textrm{Im}({\bf d}^\ast(\lambda_{j,n})) $.
\end{remark}

\begin{remark} \label{rem32}
{~}
\begin{enumerate}
\item[(i)] \ As already mentioned, the pseudo periodogram matrices  ${\bf I}^\ast(\lambda_{j,n})$  in Step I.2
are generated using   the fact that the $m$-dimensional vector of finite Fourier transforms converges towards a circular symmetric complex normal distribution. Notice that
the $ {\bf I}^\ast(\lambda_{j,n})$  generated in this step  are independent across frequencies. Therefore, the integrated periodogram  statistic  $ {\bf V}_n^\ast$ obtained in Step I.3  and based on these pseudo periodogram matrices,  is  only  able to imitate  the parts $ {\bf \Sigma_1}$ and $ {\bf \Gamma}_1$ of the limiting variance and relation matrices $ {\bf \Sigma}$ and $ {\bf \Gamma}$, respectively. This will be proven in Lemma \ref{le.BootLem2} (i). Recall that  ${\bf \Sigma}_1$ and $ {\bf \Gamma}_1$   only depend on the  spectral density matrix $ {\bf f}$ of the underlying process.
\item[(ii)] \ In Step II.1 the periodogram matrices  ${\bf I}_t(\lambda_{j,b})$ of the subsamples of length $b$  are used. In Step II.2 frequency domain residual matrices $ {\bf U}_{t}(\lambda_{j,b})  $ are defined
which are i.i.d.~resampled in Step II.3 to obtain the  pseudo periodogram matrix $ {\bf I}^+(\lambda_{j,b})$. Notice that the latter matrix  is  an  average of $k$ independent matrices
${\bf I}_{\ell}^+(\lambda_{j,b}) $ calculated  using $k$ randomly selected residual matrices $ {\bf U}_{i_\ell}(\lambda_{j,b})$ and after pre- and post-multiplying them with the square root of the
estimated spectral density matrix $ \widehat{\bf f}(\lambda_{j,b})$.   Observe that   all quantities  in Step II
are  calculated at the  Fourier frequencies $ \lambda_{j,b}=2\pi j/b$, $ j\in {\mathcal G}(b)$ corresponding to the length $b$ of the subsamples.
\item[(iii)] \ The integrated periodogram statistic   ${\bf V}_n^+$  based on $ {\bf I}^+(\lambda_{j,b})$ is  used to imitate the  missing terms $ {\bf \Sigma}_2$ and $ {\bf \Gamma}_2$
%of the asymptotic complex Gaussian distribution of interest and
 which   depend on the fourth order  moment structure  of the process. To do this  appropriately,  notice first   that the pseudo statistic $ {\bf V}_n^+$ imitates asymptotically
correct the entire covariance and relation matrices $ {\bf \Sigma}$ and $ {\bf \Gamma}$ as will be proven in Lemma \ref{le.BootLem2} (ii).  Since we use  $ {\bf V}_n^\ast$  generated in Step I to imitate   the distribution  of $ {\bf V}_n$ and  the matrices $ {\bf \Sigma}_1$ and $ {\bf \Gamma}_1$,  we have to subtract from the covariance and  relation matrix of $ {\bf V}_n^+$ the corresponding parts $ {\bf \Sigma}_{1,n}^+$ and $ {\bf \Gamma}_{1,n}^+$ so that only the desired estimators of the components
${\bf \Sigma}_2 $ and ${\bf \Gamma}_2 $ are left. This is done in Step III.1 and Step III.2.  In  particular, in Step III.1 the elements $ \sigma_{jk}^+$ and  $ c_{jk}^+$ of the matrices $ {\bf \Sigma}_{1,n}^+$ and $ {\bf \Gamma}_{1,n}^+$ are explicitly  calculated and the corresponding  matrix $ {\bf C}_n^+$ is obtained. The latter  matrix is then  subtracted from  the  matrix $ {\bf G}_n^+$, so that the obtained matrix
 $  {\bf G}_n^+-{\bf C}_n^+$   contains  the elements of the covariance and relation matrix of $ {\bf V}_n^+$ which   only depend on the fourth-order structure of the process, that is the parts $ {\bf \Sigma}_{2,n}^+ $ and $ {\bf \Gamma}_{2,n}^+$.  These are used to estimate  ${\bf \Sigma}_2 $ and $ {\bf \Gamma}_2$.
 %This is why  the difference $ {\bf G}_n^+-{\bf C}_n^+$ is calculated in this step.
 Now, adding  to this difference  the matrix ${\bf G}_n^\ast$,  leads to a  new matrix $ {\bf G}_n^\circ$ which  correctly imitates both parts  of    ${\bf \Sigma}$ and $ {\bf \Gamma}$, compare also Lemma \ref{le.BootLem2} (ii). We stress here the fact  that in $ {\bf G}_n^\circ$,   the parts  corresponding to ${\bf \Sigma}_1 $ and ${\bf \Gamma}_1 $ are contributed   by the  bootstrap procedure based on the asymptotic Gaussianity of the finite Fourier transforms,   that is by ${\bf \Sigma}_{1,n}^\ast $ and ${\bf \Gamma}_{1,n}^\ast $,
while  the parts  $ {\bf \Sigma}_2$ and ${\bf \Gamma}_2$ by the convolved bootstrap procedure, that is by $ {\bf \Sigma}_{1,n}^+$ and ${\bf \Gamma}_{1,n}^+ $.
\item[(iv)] \ In Step III.2 the matrix $ {\bf G}_n^\circ$ is used  in   equation (\ref{eq.Co})  to   appropriately rescale  the bootstrap vector $ {\bf V}_n^\ast$.
 The resulting bootstrap complex vector $ {\bf V}_n^\circ$ is   finally used  in Step III.3 to approximate  the
 distribution  of the complex vector  ${\bf V}_n$.
 \end{enumerate}
\end{remark}

\begin{remark}
%To estimate  the two parts $ {\bf \Sigma}_1$ and $ {\bf \Sigma}_2$ and ${\bf \Gamma}_1$ and $ {\bf \Gamma}_2$ of the covariance and relation
%matrices $ {\bf \Sigma}$ and $ {\bf \Gamma}$, respectively,   two different  estimators of the the spectral density matrix $ {\bf f}$ are used in the bootstrap algorithm. In particular in  StepI the estimator $ \widehat{\bf f}$ is used while in StepII the estimator $ \widetilde{\bf f}$. One can modify  the bootstrap procedure  so that  in both steps the same spectral density  estimator
% is   appleid.  One way to do this is tho set $ \widehat{\bf f}=\widetilde{\bf f}$ in StepI. However,  notice that
%the  estimator $\widehat{\bf f}$ can be  chosen by the user while the estimator $ \widetilde{\bf f}$, which is an average of periodograms  calculated over subsamples, is obtained as the expected value (in the bootstrap world) of the pseudo periogram matrices $ {\bf I}^+_\ell(\lambda_j) $ generated in Step  II.1; see also Dahlhaus (1985) for properties of the estimator $ \widetilde{\bf f}$.  Now, if one wants to use in both parts of the algorithm the same estimator $ \widehat{\bf f}$, then this can be achieved  as follows.
%Define  first the matrices
%$$ {\bf U}_{l}^+(\lambda_{l,b})  = \widetilde{\bf f}^{-1/2}(\lambda_{l,b}){\bf I}^+_{\ell}(\lambda_{l,b})  \widetilde{\bf f}^{-1/2}(\lambda_{l,b})$$
%for  $l=1,2, \ldots, [b/2]$ and $\ell =1,2, \ldots, n-b+1$.
 In the definition of the    frequency domain residual matrices ${\bf U}_{t}(\lambda_{j,b}) $ in Step II.2, the pre- and post-multiplication  with the matrix $ \widetilde{\bf f}^{-1/2}(\lambda_{j,b})$ ensures that the resampled residual matrices $ {\bf U}_{i_\ell}(\lambda_{j,b})$ satisfy
    $$ {\rm E}^*({\bf U}_{i_\ell}(\lambda_{j,b})) = \widetilde{\bf f}^{-1/2}(\lambda_{j,b}) \frac{1}{n-b+1}\sum_{t=1}^{n-b+1}{\bf I}_{t}(\lambda_{j,b})  \widetilde{\bf f}^{-1/2}(\lambda_{j,b}) =\UMa_m,$$
  where $ \UMa_m$ denotes the $m\times m $ unit matrix.
  Therefore,
\[  {\rm E}^\ast( {\bf I}^+(\lambda_{j,b}) ) = {\rm E}^\ast( {\bf I}^+_{i_\ell}(\lambda_{j,b})) =  \widehat{\bf f}^{1/2} (\lambda_{j,b})\cdot \UMa_m\cdot\widehat{\bf f}^{1/2}(\lambda_{j,b}) = \widehat{\bf f}(\lambda_{j,b}) .\]
Furthermore,    by the consistency of $ \widetilde{\bf f}(\lambda)$ as an estimator of $ {\bf f}(\lambda)$,  see the proof of Lemma 3.9, it holds true that
  for any fixed frequency $\lambda\in (0,\pi)$,
$ {\bf U}_{t}(\lambda)=\widetilde{\bf f}^{-1/2}(\lambda) {\bf I}_t(\lambda)  \widetilde{\bf f}^{-1/2}(\lambda) \stackrel{D}{\rightarrow} {\bf U}(\lambda)$, as $ b \rightarrow \infty$, where $ {\bf U}(\lambda)$  has the complex Wishart distribution of dimension $m$ and one degree of freedom, i.e., $ {\bf U}(\lambda)\sim W_m^C(1, \UMa_m)$; see Brillinger (1981), Section 4.2. Recall  that if $ {\bf U} (\lambda) \sim W_m^C(1,\UMa_m)$ then $ {\rm E}({\bf U}(\lambda))=\UMa_m$     and ${\rm  Cov}(U_{jk}(\lambda), U_{lm}(\lambda)) = \delta_{j,l}\delta_{k,m}$ for any two elements $ U_{jk}(\lambda)$ and $U_{lm} (\lambda)$ of the matrix ${\bf U}(\lambda)$.  %In Step II.1 of the bootstrap algorithm we then generate  the pseudo periodogram matrices
%$$ \widetilde{\bf I}^+_{l}(\lambda_{l,b}) =  \widehat{\bf f}^{1/2}(\lambda_{l,b})\frac{1}{k}\sum_{\ell=1}^k {\bf U}^+_{\ell}(\lambda_{l,b})  \widehat{\bf f}^{1/2}(\lambda_{l,b}).$$
%The vector   ${\bf V}_n^+$ of integrated periodograms  in StepII.2 is then replaced by
 %\[  {\bf V}^+_n = \sqrt{kb} \Big(M_{{\mathcal G}(b)}(\varphi_j,I^+_{r_j,s_j} )-M_{{\mathcal G}(b)}(\varphi_j,\widehat{f}_{r_j,s_j}) \, , \, j=1, 2, \ldots, J\Big)^\top,\]
 %and $ S_{rsuw}$ in StepIII.1 by XXX
\end{remark}

\begin{remark} The terms $ S_{rsuw}(\lambda)$ which  appear in the expressions of $ \sigma_{jk}^+$ and $c_{jk}^+$ in  Step III.1  are obtained by evaluating  the covariance expressions
  $ {\rm Cov}^\ast(I_{i_1,r_js_j}(\lambda_{\ell,b}),I_{i_1,r_ks_k}(\lambda_{\ell,b}) )$ where $ I_{i_1,rs}(\lambda_{\ell,b}) $ denotes the $ (r,s) $-th element of $ {\bf I}_{i_1}(\lambda_{\ell,b}) $. To elaborate on the parameters  they are estimating, consider  the simplified form,
  \[ \widetilde{S}_{rsuw}(\lambda) = \frac{1}{n-b+1}\sum_{t=1}^{n-b+1}\big(I_{t,rs}(\lambda)- f_{rs}(\lambda)\big) \big( I_{t,uw}(\lambda) - f_{uw}(\lambda) \big),\]
  obtained after replacing  estimated by  true quantities and where $ I_{t,rs}(\lambda)$ denotes  the $(r,s)$-th of the matrix $ {\bf I}_t(\lambda) $ defined in Step II.1. See also the proof of Lemma 3.9 (ii).
  It can then be shown  that
  \begin{equation} \label{eq.S}
  \widetilde{S}_{rsuw}(\lambda_{j,b}) = f_{rw}(\lambda_{j,b})f_{su}(-\lambda_{j,b}) + o_P(1) = f_{rw}(\lambda_{j,b})f_{us}(\lambda_{j,b}) + o_P(1).
  \end{equation}
  To see  this observe that
  \begin{align*}
  {\rm E}(\widetilde{S}_{rsuw}(\lambda_{j,b})) & =  \frac{1}{n-b+1}\sum_{t=1}^{n-b+1}{\rm E}\big((I_{t,rs}(\lambda_{j,b})- f_{rs}(\lambda_{j,b}))( \overline{I_{t,wu}(\lambda_{j,b}) - f_{wu}(\lambda_{j,b})}) \big)\\
  & = \frac{1}{n-b+1}\sum_{t=1}^{n-b+1}{\rm Cov}\big(I_{t,rs}(\lambda_{j,b}), I_{t,wu}(\lambda_{j,b}) \big) + \mathcal O(b^{-1}) \\
  & = f_{rw}(\lambda_{j,b})f_{su}(-\lambda_{j,b}) +o(1),
  \end{align*}
  where the second equality follows because $ {\rm E}(I_{t,rs}(\lambda_{j,b} )) = f_{rs}(\lambda_{j,b}) + \mathcal O(b^{-1})$ with the $ \mathcal O(b^{-1})$ term independent of $t$, and the last equality follows using   Lemma~\ref{lemmacovper}.
 By  Lemma~\ref{le.BootLem1} below,  it  can further be shown that $  {\rm Var}(\widetilde{S}_{rsuw}(\lambda_{j,b}))$ $  \rightarrow 0$ (see also the proof of Lemma~\ref{le.BootLem2}), which   justifies  expression (\ref{eq.S}).
\end{remark}

\begin{remark} Notice that if the limiting distribution of $ {\bf V}_n $ is real-valued, i.e.~if $  {\bf V} \sim \mathcal{N}(0,{\bf \Sigma},{\bf \Sigma})$, then Step III.1 and Step III.2 simplify. In particular, in this case one can set the matrices $ {\bf G}_n^\ast$ and $ {\bf G}_n^+$ as real-valued and $ J \times J$ dimensional according to $ {\bf G}_n^\ast = {\rm E}^\ast({\bf V}_n^\ast ({\bf V}_n^\ast)^\top)$ and
$ {\bf G}_n^+= {\rm E}^\ast({\bf V}_n^+ ({\bf V}_n^+)^\top)$.   Consequently, the matrix   $ {\bf C}_n^+ $ can then also be set as real-valued and $ J \times J$ dimensional according to $ {\bf C}^+_n={\bf \Sigma}^+_{1,n}$.
\end{remark}

\subsection{Smooth functions of integrated periodograms} \label{subsec_procedure_ratio}

The MFHB bootstrap procedure proposed -- appropriately modified -- also can  be applied to estimate  the distribution of  statistics which are  functions of integrated periodograms such as, for instance, sample cross-correlations. To elaborate, suppose that
\begin{equation} \label{eq.gfunction}
g=(g_1, g_2, \ldots, g_L): \C^J \rightarrow \C^L
\end{equation}
is some (smooth) function and that the statistic of interest is given by $ \sqrt{n}({\bf R}_n - {\bf R}) $ where $ {\bf R}_n = g({\bf M}_n) $ and $ {\bf R} = g({\bf M}) $ for the vector of integrated periodograms $ {\bf M}_n $ and spectral means $ {\bf M} $ given in \eqref{def_Mn} and \eqref{def_M}. Sample cross-correlations can be expressed in this way as can be seen from Example~\ref{examplesamplecorr}.\\
Our bootstrap procedure from the previous section can be adapted to approximate the distribution of $ \sqrt{n}({\bf R}_n - {\bf R}) $. We first impose some smoothness assumption on the function $ g $. To do so, we can of course interpret $ g $ as a function defined on $ \R^{2J} $ via the identification
\bee
g({\bf z})=g\begin{pmatrix}{\rm Re}({\bf z})\\
{\rm Im}({\bf z})
\end{pmatrix} \quad \quad \forall \, {\bf z} \in \C^J\,.
\eee
Splitting up real and imaginary parts also for the values of $ g $ leads to the accompanying function $ \tilde g: \R^{2J} \rightarrow \R^{2L} $ given by
\bee
\tilde g({\bf x}):= \begin{pmatrix}{\rm Re}(g({\bf x}))\\
{\rm Im}(g({\bf x}))
\end{pmatrix} \quad \quad \forall \, {\bf x} \in \R^{2J}\,.
\eee

\begin{assumption} \label{assu4}
The function $ \tilde g: \R^{2J} \rightarrow \R^{2L} $ is continuously differentiable (in the real sense) in some neighbourhood around $ ( {\rm Re}({\bf M})^\top,{\rm Im}({\bf M})^\top)^\top $ with Jacobi matrix $ {\bf J}_{\tilde g}(( {\rm Re}({\bf M})^\top,{\rm Im}({\bf M})^\top)^\top) $.
\end{assumption}

Note that -- although considering complex-valued random variables and complex functions $ g $ -- we require only real differentiability of the accompanying function $ \tilde g $. This is a much less restrictive condition than assuming complex differentiability of $ g $. We define the random vectors
\bee
\widetilde {\bf R}_n:= \begin{pmatrix}{\rm Re}({\bf R}_n)\\
{\rm Im}({\bf R}_n)
\end{pmatrix}\,, \quad \quad \widetilde {\bf R}:= \begin{pmatrix}{\rm Re}({\bf R})\\
{\rm Im}({\bf R})
\end{pmatrix} \,.
\eee
Now, applying the delta method to \eqref{CLT}, respectively \eqref{defG}, we get the limiting result
\begin{align}
\sqrt{n}(\widetilde {\bf R}_n - \widetilde {\bf R}) \stackrel{d}{\longrightarrow} {\bf J}_{\tilde g}(( {\rm Re}({\bf M})^\top,{\rm Im}({\bf M})^\top)^\top) \, \mathcal{N}_{2J}({\bf 0},{\bf G})\,.  \label{CLT_smoothfun}
\end{align}
From the continuous mapping theorem, applied for $ h(x_1,x_2):=x_1+i\cdot x_2 $, $ x_1,x_2 \in \R^{L} $, it follows
\begin{align}
\sqrt{n}({\bf R}_n - {\bf R}) \stackrel{d}{\longrightarrow} {\bf W} \sim \mathcal{N}^c_{L}({\bf 0},{\bf \Sigma_R},{\bf \Gamma_R})\,,  \label{CLT_smoothfun_g}
\end{align}
for suitable matrices $ {\bf \Sigma_R} $ and $ {\bf \Gamma_R} $ which can be obtained from \eqref{CLT_smoothfun} (the precise form of $ {\bf \Sigma_R} $ and $ {\bf \Gamma_R} $ is not needed for the ensuing bootstrap algorithm and its validity results).\\
We propose the following MFHB approximation of the distribution of $ \sqrt{n}({\bf R}_n-{\bf R})$.

\

\begin{enumerate}
\item[] {\bf Step} $\widetilde{\textbf{I}}$ \ Apply Steps I, II, III.1, and III.2 of the algorithm from Section \ref{subsec_procedure} to obtain the matrix $ {\bf G}_n^{\circ} $ and the pseudo periodogram matrices $ {\bf  I}^*(\lambda_{j,n}) $, $ j \in \mathcal{G}(n) $.
\item[] {\bf Step} $\widetilde{\textbf{II}}$ \ Let
\begin{align*}
{\bf M}^*_n &:= \Big( M_{{\mathcal G}(n)}(\varphi_j,I^\ast_{r_j,s_j} ), \, j=1, 2, \ldots, J\Big)^\top\,,\\
\widehat{\bf M}_n &:= \Big( M_{{\mathcal G}(n)}(\varphi_j,\widehat f_{r_j,s_j} ), \, j=1, 2, \ldots, J\Big)^\top\,,
\end{align*}
and
\begin{align*}
\begin{pmatrix}{\rm Re}({\bf W}^*_n)\\ {\rm Im}({\bf W}^*_n)\end{pmatrix} &:= \sqrt{n} \left[ \tilde g\begin{pmatrix}{\rm Re}({\bf M}^*_n)\\
{\rm Im}({\bf M}^*_n)
\end{pmatrix} - \tilde g\begin{pmatrix}{\rm Re}(\widehat{\bf M}_n)\\
{\rm Im}(\widehat{\bf M}_n)
\end{pmatrix} \right] \,.
\end{align*}
\item[] {\bf Step} $\widetilde{\textbf{III}}$ \ Let
\begin{align*}
\widetilde{\bf G}^*_n &:= E^*\left[\begin{pmatrix}{\rm Re}({\bf W}^*_n)\\ {\rm Im}({\bf W}^*_n)\end{pmatrix} \begin{pmatrix}{\rm Re}({\bf W}^*_n)\\ {\rm Im}({\bf W}^*_n)\end{pmatrix}^\top\right] - E^*\begin{pmatrix}{\rm Re}({\bf W}^*_n)\\ {\rm Im}({\bf W}^*_n)\end{pmatrix} \cdot E^*\begin{pmatrix}{\rm Re}({\bf W}^*_n)\\ {\rm Im}({\bf W}^*_n)\end{pmatrix}^\top
\end{align*}
and
\begin{align*}
\widetilde{\bf G}^\circ_n &:= {\bf J}_{\tilde g}\begin{pmatrix}{\rm Re}(\widehat{\bf M}_n)\\ {\rm Im}(\widehat{\bf M}_n)\end{pmatrix} \times {\bf G}^\circ_n \times {\bf J}_{\tilde g}\begin{pmatrix}{\rm Re}(\widehat{\bf M}_n)\\ {\rm Im}(\widehat{\bf M}_n)\end{pmatrix}^\top\,.
\end{align*}
\item[] {\bf Step} $\widetilde{\textbf{IV}}$ \ Calculate
\bee
\begin{pmatrix}{\rm Re}({\bf W}^\circ_n)\\ {\rm Im}({\bf W}^\circ_n)\end{pmatrix} :=  \big(\widetilde{\bf G}_{n}^\circ \big)^{1/2}\big(\widetilde{\bf G}^*_n \big)^{-1/2}
\begin{pmatrix}{\rm Re}({\bf W}^*_n)\\ {\rm Im}({\bf W}^*_n)\end{pmatrix}\,.
\eee
\item[] {\bf Step} $\widetilde{\textbf{V}}$ \ Approximate the distribution of $\sqrt{n}({\bf R}_n-{\bf R})$ by that of
\begin{align*}
{\bf W}_n^\circ & := {\rm Re}({\bf W}_n^\circ)+ i \,{\rm Im}({\bf W}_n^\circ)\,.
\end{align*}
\end{enumerate}

\vspace*{0.2cm}

\begin{remark}
{~}
\begin{enumerate}
\item[(i)] \ The algorithm above works on real and imaginary parts separately. This has the advantage that we do not have to impose complex differentiability assumptions on the function $ g $; it suffices to assume real differentiability of $ \tilde g $.
\item[(ii)] \ In Step $ \widetilde{\textrm{II}} $ the bootstrap vector $ ( {\rm Re}({\bf W}_n^*)^\top,{\rm Im}({\bf W}_n^*)^\top)^\top $ imitates the structure of
\bee
\sqrt{n}(\widetilde {\bf R}_n - \widetilde {\bf R}) = \sqrt{n} \left[ \tilde g\begin{pmatrix}{\rm Re}({\bf M}_n)\\
{\rm Im}({\bf M}_n)
\end{pmatrix} - \tilde g\begin{pmatrix}{\rm Re}({\bf M})\\
{\rm Im}({\bf M})
\end{pmatrix} \right]\,.
\eee
But as a comparison of \eqref{Th39_h3} in the proof of Theorem~\ref{th.2} with \eqref{CLT_smoothfun} shows, the limiting distribution of $ \sqrt{n}(\widetilde {\bf R}_n - \widetilde {\bf R}) $ is only partially captured by $ ( {\rm Re}({\bf W}_n^*)^\top,{\rm Im}({\bf W}_n^*)^\top)^\top $: While the factor determined by the Jacobi matrix is established properly, the matrix $ {\bf G} $ entering the covariance structure of the normal distribution is not captured. This is corrected in Steps $ \widetilde{\textrm{III}} $ and $ \widetilde{\textrm{IV}} $ where the bootstrap random vectors are first standardized via multiplication with $ \big(\widetilde{\bf G}^*_n \big)^{-1/2} $. Then the proper variance is established by $ \big(\widetilde{\bf G}^\circ_n \big)^{1/2} $ which simultaneously approximates both $ {\bf G} $ and the Jacobi matrix $ {\bf J}_{\tilde g}(( {\rm Re}({\bf M})^\top,{\rm Im}({\bf M})^\top)^\top) $.
\end{enumerate}
\end{remark}

\begin{example}
One very important statistic of interest that can be written as a smooth function of integrated periodograms is the sample cross-correlation. Examples \ref{examplesamplecorr} and \ref{examplesamplecorrcontd} show how the function $ g $ can be defined such that $ \sqrt{n}(\widehat \rho_{rs}(h)-\rho_{rs}(h)) $ takes the form $ \sqrt{n}({\bf R}_n - {\bf R}) $. The limiting variance of this expression in general takes the form indicated in \eqref{CLT_smoothfun_g}, and for this particular example it is given by the expression $ \tau^2 $ from Example \ref{examplesamplecorrcontd}. Since the function $ g $ fulfils Assumption \ref{assu4}, our bootstrap algorithm $ {\bf W}_n^\circ $ is asymptotically valid for this statistic (as is proven in Theorem \ref{th.2}), and successfully imitates the rather complicated expression $ \tau^2 $.
\end{example}

\subsection{Bootstrap validity}

We need the following consistency assumption for the spectral density estimator $ \widehat{\bf f} $ used in Step I of the bootstrap procedure in Section \ref{subsec_procedure}.

\begin{assumption} \label{assu5}
The spectral density estimator $ \widehat{\bf f} $ is Hermitian,  positive definite and satisfies
\bee
\sup_{\lambda\in[-\pi,\pi]}\|\widehat{\bf f}(\lambda)-{\bf f}(\lambda)\|_F \stackrel{P}{\rightarrow} 0\,.
\eee
%where $ \|\cdot\|_F $ denotes the Frobenius norm.
\end{assumption}

We refer to Wu and Zaffaroni (2018) for estimators of the spectral density matrix and for general classes of multivariate processes for which the above
assumption is satisfied. We next  establish the following two lemmas, the proofs of which are given in the Supplementary Material.
\vspace*{0.2cm}

\begin{lemma} \label{le.BootLem1}
Let Assumption \ref{assu1} be fulfilled. Denote by $ I_{t;rs}(\lambda) $ the $(r,s)$-th element of the periodogram matrix $ {\bf I}_t(\lambda)$  based on the subsample
$ {\bf X}(t), {\bf X}(t+1), \ldots, {\bf X}(t+b-1)$. Then,
\begin{enumerate}
\item[(i)] \ $ \sum_{\ell \in {\mathcal G}(b)} \big|\widetilde f_{rs}(\lambda_{\ell,b}) -E(I_{1;rs}(\lambda_{\ell,b})) \big| = {\mathcal O}_P(\sqrt{b^3/(n-b+1)})\,,$
\item[(ii)]  \begin{align*}  \sum_{\ell_1,\ell_2\in {\mathcal G}(b)}  & \Big|\frac{\displaystyle 1}{\displaystyle n-b+1}\sum_{t=1}^{n-b+1} I_{t;r_js_j}(\lambda_{\ell_1,b})I_{t;r_ks_k}(\lambda_{\ell_2,b}) \\
    & - E\big(I_{1;r_js_j}(\lambda_{\ell_1,b})I_{1;r_ks_k}(\lambda_{\ell_2,b}) \big) \Big| ={\mathcal O}_P(\sqrt{b^5/(n-b+1)})\,,
\end{align*}
and
  \begin{align*}  \sum_{\ell\in {\mathcal G}(b)}  & \Big|\frac{\displaystyle 1}{\displaystyle n-b+1}\sum_{t=1}^{n-b+1} I_{t;r_js_j}(\lambda_{\ell,b})I_{t;r_ks_k}(\lambda_{\ell,b}) \\
    & - E\big(I_{1;r_js_j}(\lambda_{\ell,b})I_{1;r_ks_k}(\lambda_{\ell,b}) \big) \Big| ={\mathcal O}_P(\sqrt{b^3/(n-b+1)})\,.
\end{align*}

\end{enumerate}
where the $ {\mathcal O}_P$ terms are uniformly in $r,s$, respectively $ r_j,s_j,r_k,s_k$.
\end{lemma}

\vspace*{0.2cm}

In order to formulate  our theoretical  results  dealing with the asymptotic properties of the bootstrap approximations proposed, we state the following assumption which summarizes our requirements regarding
the behavior of the subsampling parameter $b$.

\begin{assumption} \label{assu6}
$ b=b(n) \rightarrow \infty$ as $ n \rightarrow \infty$ such that $ b^3/(n-b+1) \rightarrow 0$.
\end{assumption}

\vspace*{0.2cm}

The following lemma investigates  the asymptotic properties  of the covariance and relation matrices of the random vectors $ {\bf V}^\ast_n$ and $ {\bf V}^+_n $  generated in the
MFHB procedure  and derives the
limiting distribution of  $ {\bf V}^\ast_n$.

\begin{lemma} \label{le.BootLem2}  If  Assumptions \ref{assu1}, \ref{assu2}, \ref{assu3}, \ref{assu5}  and  \ref{assu6} are satisfied, then,
\begin{enumerate}
\item[(i)] \ $ {\rm Cov}^\ast({\bf V}_n^\ast)  \stackrel{P}{\rightarrow} {\bf \Sigma}_1$ and $ E^\ast({\bf V}_n^\ast ({\bf V}_n^{\ast})^{ \top}) \stackrel{P}{\rightarrow} {\bf \Gamma}_1 $.
\item[]
\item[(ii)] \ $ {\bf G}_n^+ - {\bf C}_n^+ \stackrel{P}{\rightarrow}
\left(\begin{array}{cc}  \frac{1}{2}({\rm Re}({\bf \Sigma}_2 )+{\rm Re}({\bf \Gamma}_{2}))  &    \frac{1}{2}(-{\rm Im}({\bf \Sigma}_{2}) +{\rm Im}({\bf \Gamma}_{2}))\\
 & \\
   \frac{1}{2}({\rm Im}({\bf \Sigma}_{2}) +{\rm Im}({\bf \Gamma}_{2}))&    \frac{1}{2}({\rm Re}({\bf \Sigma}_{2}) -{\rm Re}({\bf \Gamma}_{2}))
 \end{array} \right).$
 \item[]
\item[(iii)] \ $ {\bf V}^\ast_n \stackrel{d}{\rightarrow}  {\mathcal N}_J^c({\bf 0}, {\bf \Sigma}_1, {\bf \Gamma}_1)$ in $ P $-probability.
\end{enumerate}
\end{lemma}

\vspace*{0.2cm}

Lemma~\ref{le.BootLem2} leads  to the following result which establishes consistency of the MFHB procedure in estimating the distribution and the second-order moments of the random vector $ {\bf V}_n$.

\begin{theorem}  \label{th.1}  If Assumptions \ref{assu1}, \ref{assu2}, \ref{assu3}, \ref{assu5} and \ref{assu6} are satisfied, then,
\begin{enumerate}
\item[(i)] \ $  {\rm Cov}^\ast({\bf V}_{n}^\circ) \stackrel{P}{\rightarrow} {\bf \Sigma}$, and $ E^\ast \big({\bf V}_{n}^\circ ( {\bf V}_{n}^\circ)^{\top}\big) \stackrel{P}{\rightarrow} {\bf \Gamma} $.
\item[(ii)] \ $  {\bf V}_{n}^\circ \stackrel{d}{\rightarrow}  {\bf V} $, in $ P $-probability.
\end{enumerate}
\end{theorem}

The next result   establishes  validity of the MFHB  procedure for smooth functions of integrated periodograms.

\begin{theorem}  \label{th.2}  Let  Assumptions \ref{assu1} to \ref{assu6} be  satisfied.
Then, as $n \rightarrow \infty$,
\bee
{\bf W}^\circ_n \stackrel{d}{\longrightarrow} {\bf W}
\eee
in $ P $-probability, where $ {\bf W}$ is the limiting distribution of $ \sqrt{n}({\bf R}_n-{\bf R}) $ given in \eqref{CLT_smoothfun_g}.
\end{theorem}

\section{Simulations}
\subsection{Choice of the MFHB parameters} \label{SubSim1} The practical implementation of the MFHB procedure requires the choice of two parameters. The first is the spectral density estimator $ \widehat{\f}$  and the second the subsampling parameter $b$.  Assumption 5 and Assumption 6 state our general requirements  on these parameters  focusing on the asymptotic properties they have to satisfy in order   for the proposed bootstrap procedure to be consistent.  Certainly, the choice of these parameters for a given sample size $n$ is an important venue of feature research.
In the following we discuss some rather practical  rules  on how to choose these parameters..

Regarding the spectral density estimator $\widehat{\f}$,  a variety of estimators exists which can be used in our procedure; see  Brillinger (1981).
As a simple approach, we use  in the following kernel estimators  obtained by locally averaging
the periodogram matrix over frequencies close to the frequency of interest, i.e., $\widehat{\f}(\lambda) = (nh)^{-1}\sum_{j} K((\lambda-\lambda_{j,n})/h) {\bf I}_n(\lambda_{j,n})$.  $K$ denotes the kernel function which determines  the weights assigned to the  ordinates of the  periodogram matrix, while  $h$ is the bandwidth that  controls the  number of periodogram ordinates effectively taken into account in order to obtain the kernel estimator $\widehat{\f}$.
To select the parameter $h$  in practice, cross-validation type approaches have been proposed and investigated in the literature which also can be applied in our setting; see Robinson (1991) for details.

For the subsampling parameter $b$, Assumption 6 solely states the required conditions on   the rate at which this parameter has to  increase to infinity with respect  to  the sample size $n$  in order to ensure consistency of the MFHB  procedure. Our simulation  experience with the choice of this parameter  shows that the  results obtained are not very sensitive with respect to the choice of $b$, provided that this parameter is not chosen too small. This motivates   the  suggestion of the  following
practical rule for selecting  $b$. Select this parameter as the smallest   integer which is larger  or equal to $ 3\cdot n^{0.30}$. This rule satisfies the requirements of Assumption 6 and at the same time delivers a value of $b$ which is large enough for the MFHB procedure to perform well in practice.

Notice that the numerical results   presented in the next  section are reported for
different combinations of bandwidth and block size parameters, $ h $ and $b$. On the one hand this avoids a further  increase of  the computational burden caused by a   cross-validation type choice of the bandwidth $h$. On the other hand it allows us to investigate the  sensitivity of the bootstrap estimates with respect to  different choices of the parameters involved.

\subsection{Numerical results}
In this section we investigate the finite sample performance of the MFHB procedure and compare it with that of the time domain moving block bootstrap (MBB). Note that, in view of our discussion in the Introduction, popular time domain bootstrap methods other than the MBB (and its variations) can not be considered as competitors for our MFHB since these procedures are asymptotically non-valid for a large class of statistics  in the context of multivariate linear and non-linear time series. We consider time series of length  $ n=100$ stemming from the following
 two bivariate processes considered in Tsay (2014).  The first is a  VAR(1) process driven by i.i.d. innovations, i.e.,
\[\hspace*{-2.5cm} \mbox{Model I:} \ \ \ \  {\bf X}(t) = {\bf  \Phi}{\bf X}(t-1)  + {\bf e}(t), \ \  {\bf \Phi}= \left(\begin{array}{rr} 0.8 & 0.4 \\ -0.3 & 0.6 \end{array} \right),   \]
where $ {\bf e}_t \sim {\mathcal N}_2({\bf 0}, {\bf S}_{\bf e})$ and $  {\bf S}_{\bf e}=(\sigma_{i,j})_{i,i=1,2}$, with $ \sigma_{1,1}=2.0$, $\sigma_{1,2}=0.5$ and $ \sigma_{2,2}=1$.
The second model  is a bivariate  VARMA(2,1) process the  innovations of which follow a  bivariate GARCH-type  process, that is,
\[\hspace*{-1.5cm}  \mbox{Model II:} \ \ \ \ {\bf X}(t) = {\bf  \Phi}_1{\bf X}(t-1) +   {\bf \Phi}_2{\bf X}(t-2) + {\bf \Theta} {\bf u}(t-1) + {\bf u}(t), \]
with parameter matrices  ${\bf \Phi}_1= \left(\begin{array}{rr} 0.816 & -0.623 \\ -1.116 & 1.074 \end{array} \right)$,
\[
{\bf \Phi}_2 = \left(\begin{array}{rr} -0.643 & 0.592 \\ 0.615 & -0.133 \end{array} \right) \  \mbox{and} \ \
{\bf \Theta}= \left(\begin{array}{rr} 0 & -1.248 \\ -0.801 & 0 \end{array} \right).\]
Furthermore, the innovations $ {\bf u}(t)$ are generated as  $  {\bf u}(t)={\bf S}_{t}^{1/2}{\bf e}(t)$, where  the $ {\bf e}(t)$'s  are  i.i.d.~${\mathcal N}_2({\bf 0}, {\bf Id}_2)$ distributed, and  the volatility matrix $ {\bf S}_{t}$  evolves according to a  BEKK(1,1) model,  i.e.~$ {\bf S}_{t}={\bf A}_0{\bf A}^\top_0 + {\bf A}_1{\bf u}(t-1){\bf u}(t-1)^\top {\bf A}_1^\top + {\bf B}_1 {\bf S}_{t-1} {\bf B}_1^\top$, where
\[{\bf A}_0= \left(\begin{array}{rr} 0.01 & 0 \\ 0 & 0.01 \end{array} \right), \ {\bf A}_1= \left(\begin{array}{rr} 0.15 & 0.20 \\ 0.06 & 0.40 \end{array} \right) \ \mbox{and} \
{\bf B}_1= \left(\begin{array}{rr} 0.9 & 0 \\ 0 & 0.9 \end{array} \right).\]

Observe  that for the  VAR(1) model
 with  Gaussian innovations the fourth-order cumulant spectral densities  $f_{rsvw}$ equal zero, so the expressions of the covariance  and relation matrices   $ {\bf \Sigma} $ and $ {\bf \Gamma}$  simplify. This is not the case  for  the VARMA(2,1) process which  is nonlinear due
to the BEKK(1,1)  generated  innovations. The parameter matrices of this BEKK(1,1) model are  very  close to those of the same  model fitted to the IBM stock and S\&P composite index in Tsay (2014), p. 418, Table 7.3. See also Francq and Zako\"ian (2016), Section 6,  for  a similar parametrization.

We consider the problem of estimating the standard deviation of the cross-correlation estimates $ \widehat{\rho}(h)$ for the values $ h=-1, 0,+1$.  Recall that the estimators  considered
also can be written as functions of integrated periodograms; see Example 2.3.  We have
generated 10,000 replications of both models in order to estimate the exact standard deviation of the sample cross-correlations  considered. As already mentioned, the   spectral density estimator $ \widehat{\f}$ used in the MFHB procedure is  a kernel estimator obtained via smoothing the periodogram matrix with   bandwidth $h$ and  using the Bartlett-Priestley kernel;  see Priestley (1981).
Furthermore,  and in order  to see the sensitivity  of the bootstrap methods compared
with respect to the choice of the bootstrap parameters, the  MFHB procedure has been applied using different choices of the bandwidth $h$ and of the subsampling parameter $b$. The same values of
$b$ have also been used as block sizes in the MBB procedure.  We also report results for  $b=12$ which is the value of the subsampling parameter  selected according to the rule proposed in Section \ref{SubSim1}.

Table 1  presents the results for both models considered. The results reported in this table are   based on $R=500$ repetitions  where  $ B=300$ bootstrap replications have been applied for each repetition.  As this table shows, the MFHB procedure performs  quite well for both models considered  and outperforms  the MBB. In particular,  comparing
the performance  of both bootstrap procedures for the same subsampling parameter, respectively block size $b$, the MFHB  mean square errors
 are in almost all cases considered, and independent of the choice of $h$, lower than those of the MBB procedure. Furthermore, the MFHB  estimates seem to be  less sensitive with respect to the choice of the subsampling parameter $b$ than
 the MBB procedure is with respect to the choice of the block size $b$.

\begin{small}
\begin{table}[ht]
\label{table1}
%\begin{center}
\hspace*{-1.2cm}
\begin{tabular}{llllcllcllc}
\hline
   & & &  & & & & & &  & \\
   & &$\widehat{\rho}(-1)$ & &  & $\widehat{\rho}(0)$&   &  &$\widehat{\rho}(+1) $& &  \\
   &  & Mean& Std & MSE$\times$10& Mean & Std &  MSE$\times$10 &Mean & Std &  MSE$\times$10\\
 \hline
   & &  &  & & & & & &  & \\
%    & &  &  & & & & & &  & \\
 & & & &   \multicolumn{4}{c}{\normalsize{ MODEL I }}   & &  \\
    &  & \multicolumn{3}{l}{Est. Exact:  0.766} &\multicolumn{3}{l}{Est. Exact:  0.992} &\multicolumn{3}{l}{Est. Exact:  1.131}  \\
  {\normalsize MFHB}  & & &  & & & & & &  & \\
 h=0.10 & b=6 &  0.794 &0.123 & 0.158 &1.014 &0.129 & 0.176&  1.142& 0.167 & 0.283 \\
                     & b=8 &0.788 & 0.130 & 0.172 & 1.024 &0.140 &0.206 & 1.153& 0.188 & 0.358\\
                      & b=10 &0.787 & 0.129 &0.168 & 1.011 &0.146 &0.214 & 1.143& 0.192 & 0.368\\
                          & b=12 &0.788& 0.136 &0.188 & 1.000 &0.160 &0.256 & 1.130& 0.192 & 0.367\\
                              & b=16 &0.771 & 0.142 &0.203 & 0.958 &0.171 &0.300 & 1.104& 0.212 & 0.456\\
            & & &  & & & & & &  & \\
           h=0.12 & b=6 &0.824 & 0.124 &0.187 & 1.027 & 0.140 & 0.207 & 1.159& 0.172 & 0.308\\
                     & b=8 &0.828 & 0.122 &0.178 & 1.045 & 0.126 & 0.181 & 1.171& 0.170 & 0.300\\
                     & b=10 &0.821 & 0.117 & 0.162 & 1.030 & 0.129 & 0.181 & 1.152& 0.171 & 0.297\\
                      & b=12 &0.815& 0.127 &0.179 & 1.023 &0.147 &0.226 & 1.156& 0.170 & 0.295\\
                              & b=16 &0.799 & 0.145 &0.243 & 1.005 &0.167 &0.283 & 1.140& 0.191 & 0.365\\
            & & &  & & & & & &  & \\
            {\normalsize MBB} & & &  & & & & & &  & \\
        & b=6 &0.906 & 0.112 & 0.323 & 1.016 &0.162& 0.273 & 1.092& 0.152 & 0.248\\
              & b=8 & 0.849 & 0.122 & 0.201 & 0.977 &0.181 &0.332 & 1.055& 0.185 & 0.406\\
          & b=10 & 0.832& 0.143 & 0.237 & 0.955 & 0.198 & 0.414 & 1.048& 0.199 & 0.475\\
           & b=12 &0.795& 0.145 &0.217 & 0.933 &0.208 &0.477 & 1.020& 0.205 & 0.548\\
           & b=16 &0.750 & 0.164 &0.271 & 0.875 &0.208 &0.559 & 0.966& 0.217 & 0.743\\
  & & &  & & & & & &  & \\
 & & & &   \multicolumn{4}{c}{\normalsize{ MODEL II }}   & &  \\
  %  &  & \multicolumn{3}{l}{Est. Exact:  1.217} &\multicolumn{3}{l}{Est. Exact:  1.135} &\multicolumn{3}{l}{Est. Exact:  1.154}  \\
    &  & \multicolumn{3}{l}{Est. Exact:  1.154} &\multicolumn{3}{l}{Est. Exact:  1.135} &\multicolumn{3}{l}{Est. Exact:  1.217}  \\
  %\hline
  % & & &  & & & & & &  & \\
  {\normalsize MFHB}  & & &  & & & & & &  & \\
 h=0.10 & b=6 &  1.164 &0.232 & 0.570 &1.160 &0.166 & 0.300  &1.195& 0.196 & 0.424 \\
                     & b=8 &1.153 & 0.263 &0.620 & 1.168 &0.168 &0.301 & 1.196& 0.195 & 0.409\\
                      & b=10 &1.140 & 0.235 &0.633 & 1.148&0.181 &0.331 & 1.193& 0.183 & 0.353\\
                       & b=12 &1.113& 0.240 &0.785 & 1.115 &0.205 &0.428 & 1.193& 0.216 & 0.491\\
                              & b=16 &1.113 & 0.248 &0.742 & 1.084 &0.229 &0.536 & 1.144& 0.245 & 0.601\\
            & & &  & & & & & &  & \\
           h=0.12 & b=6 &1.167 & 0.224 & 0.544 & 1.195 &0.173 &0.337 & 1.222& 0.195 & 0.431\\
                     & b=8 &1.158 & 0.222 & 0.547 & 1.193 & 0.160 & 0.311 & 1.216& 0.185 & 0.399\\
                     & b=10 &1.184 & 0.230 & 0.538 & 1.200 &0.176 & 0.357 & 1.227& 0.186 & 0.387\\
                      & b=12 &1.149& 0.227 &0.578 & 1.159 &0.194 &0.394 & 1.221& 0.203 & 0.498\\
                              & b=16 &1.134 & 0.235 &0.661 & 1.111 &0.208 &0.437 & 1.168& 0.232 & 0.538\\
            & & &  & & & & & &  & \\
            {\normalsize MBB} & & &  & & & & & &  & \\
        & b=6 &1.176& 0.208 & 0.454 & 1.229 & 0.255 &0.784 & 1.191& 0.221 & 0.524\\
              & b=8 &1.157 & 0.245 &0.655 & 1.189 &0.273 & 0.782 & 1.187& 0.246 & 0.623\\
          & b=10 &1.096 & 0.235 &0.730 & 1.131 &0.268 &0.718 & 1.162& 0.263 & 0.695\\
           & b=12 &1.048& 0.238 &0.999 & 1.090 &0.271 &0.766 & 1.116& 0.269 & 0.735\\
                              & b=16 &1.024 & 0.245 &1.002 & 1.057 &0.289 &0.870 & 1.080& 0.290 & 0.888\\
  & & &  & & & & & &  & \\
\hline
%\end{center}
\end{tabular}
%\end{center}
{\normalsize  {\rm TABLE 1: \ Bootstrap estimates of the standard deviation of the  sample cross-correlations $ \widehat{\rho}(h)$ for lags $ h \in \{-1,0,+1\}$  for time series of length $n=100$ stemming  from Model I and Model II.
MFHB refers to the estimates of the multivariate frequency domain hybrid bootstrap and MBB to those of the moving block bootstrap.}}
\end{table}
\end{small}

\section{Appendix: Proofs}

$\left.\right.$\\

{\sc Proof of Lemma~\ref{le.BootLem2} $(i)$:}
The $ (j,k) $-th entry of the covariance matrix $ {\rm Cov}^\ast({\bf V}_{n}^\ast) $ is given by
\bee
{\rm Cov}^\ast(V_{n,j}^\ast,V_{n,k}^\ast) &=& {\rm Cov}^\ast \left( \frac{2\pi}{\sqrt n} \sum_{l_1 \in \mathcal G(n)} \varphi_j(\lambda_{l_1,n}) \big( I^\ast_{r_js_j}(\lambda_{l_1,n}) - \widehat f_{r_js_j}(\lambda_{l_1,n}) \big), \right. \\
& & \quad \quad \quad \left. \frac{2\pi}{\sqrt n} \sum_{l_2 \in \mathcal G(n)} \varphi_k(\lambda_{l_2,n}) \big( I^\ast_{r_ks_k}(\lambda_{l_2,n}) - \widehat f_{r_ks_k}(\lambda_{l_2,n}) \big) \right)
\eee
which equals
\be
\quad \quad \frac{4\pi^2}{n} \sum_{l_1,l_2 \in \mathcal G(n)} \varphi_j(\lambda_{l_1,n}) \overline{\varphi_k(\lambda_{l_2,n})} \, {\rm Cov}^\ast\big( I^\ast_{r_js_j}(\lambda_{l_1,n}), I^\ast_{r_ks_k}(\lambda_{l_2,n}) \big)\,. \label{proofh1}
\ee
The last expression can be decomposed, using $ \lambda_{-l,n}=-\lambda_{l,n} $ and $ I^\ast_{rs}(-\lambda_{l,n})=I^\ast_{sr}(\lambda_{l,n}) $, into
\begin{align}
\frac{4\pi^2}{n} \sum_{l_1,l_2=1}^{N} &\left\{ \varphi_j(\lambda_{l_1,n}) \overline{\varphi_k(\lambda_{l_2,n})} \, {\rm Cov}^\ast\big( I^\ast_{r_js_j}(\lambda_{l_1,n}), I^\ast_{r_ks_k}(\lambda_{l_2,n}) \big) \right. \label{proofh2}\\
&+  \varphi_j(-\lambda_{l_1,n}) \overline{\varphi_k(-\lambda_{l_2,n})} \, {\rm Cov}^\ast\big( I^\ast_{s_jr_j}(\lambda_{l_1,n}), I^\ast_{s_kr_k}(\lambda_{l_2,n}) \big) \nonumber\\
&+  \varphi_j(-\lambda_{l_1,n}) \overline{\varphi_k(\lambda_{l_2,n})} \, {\rm Cov}^\ast\big( I^\ast_{s_jr_j}(\lambda_{l_1,n}), I^\ast_{r_ks_k}(\lambda_{l_2,n}) \big) \nonumber\\
&\left.+  \varphi_j(\lambda_{l_1,n}) \overline{\varphi_k(-\lambda_{l_2,n})} \, {\rm Cov}^\ast\big( I^\ast_{r_js_j}(\lambda_{l_1,n}), I^\ast_{s_kr_k}(\lambda_{l_2,n}) \big) \right\}\,. \nonumber
\end{align}
For the first of the four similar summands in this expression we can calculate the following, where we used that $ {\bf I}^\ast(\lambda_{l_1,n}) $ and $ {\bf I}^\ast(\lambda_{l_2,n}) $ are independent for $ l_1\neq l_2 $ and $ 1 \leq l_1,l_2 \leq N $, and that $ {\bf I}^\ast(\lambda_{l_1,n}) $ has a complex Wishart $ W^C_m(1,{\bf \widehat f}(\lambda_{l_1,n}) )$ distribution (the covariance structure of which can be obtained from, e.g., Brillinger  (1981), Section 4.2):
\begin{align*}
& \frac{4\pi^2}{n} \sum_{l_1,l_2=1}^{N} \varphi_j(\lambda_{l_1,n}) \overline{\varphi_k(\lambda_{l_2,n})} \, {\rm Cov}^\ast\big( I^\ast_{r_js_j}(\lambda_{l_1,n}), I^\ast_{r_ks_k}(\lambda_{l_2,n}) \big) \\
=\, & \frac{4\pi^2}{n} \sum_{l=1}^{N} \varphi_j(\lambda_{l,n}) \overline{\varphi_k(\lambda_{l,n})} \, {\rm Cov}^\ast\big( I^\ast_{r_js_j}(\lambda_{l,n}), I^\ast_{r_ks_k}(\lambda_{l,n}) \big) \\
=\, & \frac{4\pi^2}{n} \sum_{l=1}^{N} \varphi_j(\lambda_{l,n}) \overline{\varphi_k(\lambda_{l,n})} \, \widehat f_{r_jr_k}(\lambda_{l,n}) \, \overline{\widehat f_{s_js_k}(\lambda_{l,n})} \\
=\, & \frac{4\pi^2}{n} \sum_{l=1}^{N} \varphi_j(\lambda_{l,n}) \overline{\varphi_k(\lambda_{l,n})} \, f_{r_jr_k}(\lambda_{l,n}) \, f_{s_js_k}(-\lambda_{l,n}) \\
& + \frac{4\pi^2}{n} \sum_{l=1}^{N} \varphi_j(\lambda_{l,n}) \overline{\varphi_k(\lambda_{l,n})} \, \big(\widehat f_{r_jr_k}(\lambda_{l,n}) \, \widehat f_{s_js_k}(-\lambda_{l,n}) - f_{r_jr_k}(\lambda_{l,n}) \, f_{s_js_k}(-\lambda_{l,n}) \big)\,.
\end{align*}
The second summand on the last right-hand side vanishes asymptotically due to Assumption \ref{assu5} and $ N=\mathcal O(n) $. Since the first summand is a Riemann sum the last right-hand side converges in probability to
\bee
2\pi \int_0^\pi \varphi_j(\lambda) \overline{\varphi_k(\lambda)} \, f_{r_jr_k}(\lambda) f_{s_js_k}(-\lambda) \, d\lambda\,.
\eee
With analogous calculations the other three summands in \eqref{proofh2} yield
\bee
& 2\pi \int_0^\pi \varphi_j(-\lambda) \overline{\varphi_k(-\lambda)} \, f_{s_js_k}(\lambda) f_{r_jr_k}(-\lambda) \, d\lambda \\
+ & 2\pi \int_0^\pi \varphi_j(-\lambda) \overline{\varphi_k(\lambda)} \, f_{s_jr_k}(\lambda) f_{r_js_k}(-\lambda) \, d\lambda \\
+ & 2\pi \int_0^\pi \varphi_j(\lambda) \overline{\varphi_k(-\lambda)} \, f_{r_js_k}(\lambda) f_{s_jr_k}(-\lambda) \, d\lambda\,.
\eee
Substituting $ \lambda $ with $ -\lambda $ in the first and second of these three terms shows that the limit in probability of \eqref{proofh2} is given by $ \Sigma_{1;jk} $.\\
As for the relation matrix, observe that the $ (j,k) $-th entry of $ E^\ast({\bf V}_{n}^\ast {\bf V}_{n}^{\ast \top}) $ is given by
\bee
{\rm Cov}^\ast(V_{n,j}^\ast,\overline{V_{n,k}^\ast}) = \frac{4\pi^2}{n} \sum_{l_1,l_2 \in \mathcal G(n)} \varphi_j(\lambda_{l_1,n}) \varphi_k(-\lambda_{l_2,n}) \, {\rm Cov}^\ast\big( I^\ast_{r_js_j}(\lambda_{l_1,n}), I^\ast_{r_ks_k}(\lambda_{l_2,n}) \big)\,,
\eee
which is \eqref{proofh1} if one replaces $ \varphi_k(\cdot) $ with $ \overline{\varphi_k(-\,\cdot)} $. Therefore, one can follow along the lines of the calculation for $ \Sigma_{1;jk} $ above to see that the $ (j,k) $-th entry of $ E^\ast({\bf V}_{n}^\ast {\bf V}_{n}^{\ast \top}) $ converges in probability to $ \Gamma_{1;jk} $. $ \hfill \square $

\

{\sc Proof of Lemma~\ref{le.BootLem2} $(ii)$:} \ We first establish  the uniform (over  the Fourier frequencies) consistency of $ \widetilde{\f}^{-1/2} $ and $ \widehat{\f}^{1/2}$ as estimators of $ \f^{-1/2}$ and $ \f^{1/2}$, respectively.
We first show that
\begin{equation}\label{eq.Lemma3.9.1}
\max_{\ell \in  {\mathcal G}(b)} \| \widetilde{\f}^{-1}(\lambda_{\ell,b}) - \f^{-1}(\lambda_{\ell,b})\|_F \stackrel{P}{\rightarrow} 0.
\end{equation}
For this notice first  that
\begin{equation}\label{eq.Lemma3.9.1b}
\max_{\ell \in  {\mathcal G}(b)} \| \widetilde{\f}(\lambda_{\ell,b}) - \f(\lambda_{\ell,b})\|_F \stackrel{P}{\rightarrow} 0.
\end{equation}
This follows since $ E(I_{1;rs}(\lambda_{\ell,b}) )= f_{rs}(\lambda_{\ell,b}) + \mathcal O(b^{-1})$ uniformly in $r,s$ and
\begin{align*}
\max_{\ell \in  {\mathcal G}(b)} \| \widetilde{\f}(\lambda_{\ell,b}) - \f(\lambda_{\ell,b})\|_F & \leq  \max_{\ell \in  {\mathcal G}(b)} \| \widetilde{\f}(\lambda_{\ell,b}) - E\widetilde{\f}(\lambda_{\ell,b})\|_F\\
&  \ \ \ \ \ \  + \max_{\ell \in  {\mathcal G}(b)} \| E\widetilde{\f}(\lambda_{\ell,b}) - \f(\lambda_{\ell,b})\|_F\\
& =  \max_{\ell \in  {\mathcal G}(b)} \Big\{\sum_{r=1}^m\sum_{s=1}^m \big| \widetilde{f}_{rs}(\lambda_{\ell,b}) -  E \widetilde{f}_{rs}(\lambda_{\ell,b})  \big|^2 \Big\}^{1/2}  \\
 & \ \    + \max_{\ell \in  {\mathcal G}(b)} \Big\{\sum_{r=1}^m\sum_{s=1}^m \big| E(I_{1;rs}(\lambda_{\ell,b})) -  f_{rs}(\lambda_{\ell,b})  \big|^2 \Big\}^{1/2}\\
 & \leq    \sum_{r=1}^m\sum_{s=1}^m \sum_{\ell\in{\mathcal G}(b)}\big| \widetilde{f}_{rs}(\lambda_{\ell,b}) -  E \widetilde{f}_{rs}(\lambda_{\ell,b})  \big| + \mathcal O(b^{-1})\\
 & = \mathcal O_P(\sqrt{b^3/(n-b+1)}) + \mathcal O(b^{-1}) \rightarrow 0,
\end{align*}
where the last convergence holds true because of Lemma 3.8 (i) and Assumption 6.
Now to see  (\ref{eq.Lemma3.9.1})  notice that $ \max_{\ell\in{\mathcal G}(b)}\|\f^{-1}(\lambda_{\ell,b})\|^2_F =\mathcal O(1)$ and that
 \begin{align} \label{eq.hilf1}
  \max_{\ell \in  {\mathcal G}(b)} \| \widetilde{\f}^{-1} & (\lambda_{\ell,b})     -
 \f^{-1}(\lambda_{\ell,b})\|_F  =    \max_{\ell \in  {\mathcal G}(b)} \| \widetilde{\f}^{-1}(\lambda_{\ell,b}) \big( \f(\lambda_{\ell,b}) - \widetilde{\f}(\lambda_{\ell,b})\big)
\f^{-1}(\lambda_{\ell,b})\|_F \nonumber  \\
 & \leq    \max_{\ell \in  {\mathcal G}(b)} \| \widetilde{\f}^{-1}(\lambda_{\ell,b})\|_F   \max_{\ell \in  {\mathcal G}(b)} \| \widetilde{\f}(\lambda_{\ell,b}) - \f(\lambda_{\ell,b})\|_F  \max_{\ell \in  {\mathcal G}(b)} \| \f^{-1}(\lambda_{\ell,b})\|_F
 \nonumber \\
 & \leq \big(     \max_{\ell \in  {\mathcal G}(b)} \| \f^{-1}(\lambda_{\ell,b})\|_F +    \max_{\ell \in  {\mathcal G}(b)} \| \widetilde{\f}^{-1}(\lambda_{\ell,b}) - \f^{-1}(\lambda_{\ell,b})\|_F\big) \nonumber  \\
 & \ \ \ \times     \max_{\ell \in  {\mathcal G}(b)} \| \widetilde{\f}(\lambda_{\ell,b}) - \f(\lambda_{\ell,b})\|_F  \max_{\ell \in  {\mathcal G}(b)} \| \f^{-1}(\lambda_{\ell,b})\|_F
 \end{align}
 By (\ref{eq.Lemma3.9.1b})  and for $n$ large enough such that
 $$ \max_{\ell \in {\mathcal G}(b)}\|\widetilde{\f}(\lambda_{\ell,b})-\f(\lambda_{\ell,b})\|_F \max_{\ell \in {\mathcal G}(b)}\|\f^{-1}(\lambda_{\ell,b})\|_F< 1,$$
 expression  (\ref{eq.hilf1})
  leads to
 \begin{align*}
  \max_{\ell \in  {\mathcal G}(b)} \| \widetilde{\f}^{-1}(\lambda_{\ell,b})  & -
 \f^{-1}(\lambda_{\ell,b})\|_F \\
 &  \leq \frac{\displaystyle \max_{\ell \in {\mathcal G}(b)}\|\f^{-1}(\lambda_{\ell,b})\|^2_F \max_{\ell \in {\mathcal G}(b)}\|\widetilde{\f}(\lambda_{\ell,b})-\f(\lambda_{\ell,b})\|_F }{\displaystyle 1- \max_{\ell \in {\mathcal G}(b)}\|\f^{-1}(\lambda_{\ell,b})\|_F \max_{\ell \in {\mathcal G}(b)}\|\widetilde{\f}(\lambda_{\ell,b})-\f(\lambda_{\ell,b})\|_F }
  \stackrel{P}{\rightarrow} 0.
 \end{align*}

 Assumption \ref{assu5} together with (\ref{eq.Lemma3.9.1}) implies that
\begin{equation}
\label{eq.Lemma3.9.2}
\max_{\ell \in  {\mathcal G}(b)} \| \widehat{\f}^{1/2}(\lambda_{\ell,b}) - \f^{1/2}(\lambda_{\ell,b})\|_F \stackrel{P}{\rightarrow} 0
\end{equation}
and
\begin{equation}
\label{eq.Lemma3.9.3}
\max_{\ell \in  {\mathcal G}(b)} \| \widetilde{\f}^{-1/2}(\lambda_{\ell,b}) - \f^{-1/2}(\lambda_{\ell,b})\|_F \stackrel{P}{\rightarrow} 0.
\end{equation}
To see (\ref{eq.Lemma3.9.2}) notice that  since $ \widehat{\f}$ is positive definite, we get by Assumption 2 and equation (1.3) in Schmitt (1992),  that
\[ \max_{\ell \in  {\mathcal G}(b)} \| \widehat{\f}^{1/2}(\lambda_{\ell,b}) - \f^{1/2}(\lambda_{\ell,b})\|_F \leq \frac{1}{\sqrt{\delta}}  \max_{\ell \in  {\mathcal G}(b)}\| \widehat{\f}(\lambda_{\ell,b}) - \f(\lambda_{\ell,b})\|_F.\]
For (\ref{eq.Lemma3.9.3}) we get by the same arguments  as above and since $ {\bf A}^{-1/2} = ({\bf A}^{-1})^{1/2}$,  the bound
 \[ \max_{\ell \in  {\mathcal G}(b)} \| \widetilde{\f}^{-1/2}(\lambda_{\ell,b}) - \f^{-1/2}(\lambda_{\ell,b})\|_F \leq \sqrt{\sigma_{\max}}   \max_{\ell \in  {\mathcal G}(b)} \| \widetilde{\f}^{-1}(\lambda_{\ell,b}) -
 \f^{-1}(\lambda_{\ell,b})\|_F,\]
 where $ \sigma_{\max} = \max_{\lambda\in [0,\pi]} \max\sigma(f(\lambda))$.  Equation  (\ref{eq.Lemma3.9.3}) follows then  by (\ref{eq.Lemma3.9.1}).

Assertions (\ref{eq.Lemma3.9.2}) and  (\ref{eq.Lemma3.9.3}) imply that
\begin{equation}
\label{eq.Root1}
\max_{\ell \in  {\mathcal G}(b)} \| \widehat{\f}^{1/2}(\lambda_{\ell,b}) \widetilde{\f}^{-1/2}(\lambda_{\ell,b}) - \UMa_m\|_F \stackrel{P}{\rightarrow} 0
\end{equation}
and
\begin{equation}
\label{eq.Root2}
\max_{\ell \in  {\mathcal G}(b)} \| \widetilde{\f}^{-1/2}(\lambda_{\ell,b}) \widehat{\f}^{1/2}(\lambda_{\ell,b}) - \UMa_m\|_F \stackrel{P}{\rightarrow} 0.
\end{equation}
Using (\ref{eq.Root1}) and (\ref{eq.Root2}) it follows by straightforward calculations that
\[ \textrm{Cov}^\ast\big(V_{n,j}^+,V_{n,k}^+\big) = \textrm{Cov}^\ast\big(\widetilde{V}_{n,j},\widetilde{V}_{n,k}\big) + o_P(1),\]
where
\[ \widetilde{V}_{n,j} = \sqrt{kb} \frac{2\pi}{b}\sum_{\ell \in {\mathcal G}(b)} \varphi_j(\lambda_{\ell,b}) \big(\widetilde{I}_{r_js_j}(\lambda_{\ell,b})- \widetilde{f}_{r_js_j}(\lambda_{\ell,b}) \big)\]
and $ \widetilde{I}_{rs}(\lambda_{\ell,b})$ denotes the (r,s)-th element of the matrix $ k^{-1}\sum_{m=1}^k{\bf I}_{i_m}(\lambda_{\ell,b})$. Let
$ I_{i_1,rs}(\lambda_{\ell,b})$ denote the $(r,s)$-th element of $ {\bf I}_{i_1}(\lambda_{\ell,b}) $. We then have
\begin{align*}
\textrm{Cov}^\ast(\widetilde{V}_{n,j}, & \widetilde{V}_{n,k})  = \frac{4\pi^2}{b}\sum_{\ell_1, \ell_2\in {\mathcal G}(b)}  \varphi_j(\lambda_{\ell_1,b})\overline{\varphi_k(\lambda_{\ell_2,b})} \textrm{Cov}^\ast(I_{i_1,r_js_j}(\lambda_{\ell_1,b}), I_{i_1,r_ks_k}(\lambda_{\ell_2,b}))\\
& =\frac{4\pi^2}{b}\sum_{\ell\in {\mathcal G}(b)} \varphi_j(\lambda_{\ell,b})\overline{\varphi_k(\lambda_{\ell,b})} \textrm{Cov}^\ast(I_{i_1,r_js_j}(\lambda_{\ell,b}), I_{i_1,r_ks_k}(\lambda_{\ell,b}))\\
& \ \ + \frac{4\pi^2}{b}\sum_{\ell\in {\mathcal G}(b)} \varphi_j(\lambda_{\ell,b})\overline{\varphi_k(-\lambda_{\ell,b})} \textrm{Cov}^\ast(I_{i_1,r_js_j}(\lambda_{\ell,b}), I_{i_1,r_ks_k}(-\lambda_{\ell,b}))\\
& \ \ + \frac{4\pi^2}{b}\sum_{|\ell_1|\neq |\ell_2| \in {\mathcal G}(b)} \varphi_j(\lambda_{\ell_1,b})\overline{\varphi_k(\lambda_{\ell_2,b})}\textrm{Cov}^\ast(I_{i_1,r_js_j}(\lambda_{\ell_1,b}), I_{i_1,r_ks_k}(\lambda_{\ell_2,b}))\\
& \stackrel{P}{\rightarrow} \Sigma_{1;jk} + \Sigma_{2;jk},
\end{align*}
using
\begin{align*}
 \textrm{Cov}^\ast(& I_{i_1,r_js_j}(\lambda_{\ell_1,b}), I_{i_1,r_ks_k}(\lambda_{\ell_2,b})) \\
 & =\frac{1}{n-b+1}\sum_{t=1}^{n-b+1}\big( I_{t,r_js_j}(\lambda_{\ell_1,b})-\widetilde{f}_{r_js_j}(\lambda_{\ell_1,b})\big)
\overline{\big( I_{t,r_ks_k}(\lambda_{\ell_2,b})-\widetilde{f}_{r_ks_k}(\lambda_{\ell_2,b})\big)},
\end{align*}
and Lemma~\ref{le.BootLem1}.
By the same arguments it follows  that
\[ E^\ast\big(V_{n,j}^+V_{n,k}^+\big) = E^\ast(\widetilde{V}_{n,j}\widetilde{V}_{n,k}\big) + o_P(1),\]
and that
\[  E^\ast(\widetilde{V}_{n,j}\widetilde{V}_{n,k}\big) \stackrel{P}{\rightarrow} \Gamma_{1;jk} + \Gamma_{2;jk}.\]
The above  results show that  $ {\bf G}_n^+ \stackrel{P}{\rightarrow} {\bf G}$. To conclude the proof we have to show that $ {\bf \Sigma}_{1,n}^+ \stackrel{P}{\rightarrow} {\bf \Sigma_1}$ and that
$ {\bf \Gamma}^+_{1,n} \stackrel{P}{\rightarrow} {\bf \Gamma}_{1}$.  Using (\ref{eq.Root1}) and (\ref{eq.Root2})  again, it follows  that
\begin{align*}
    \sigma^+_{jk} & = \frac{4\pi^2}{b} \sum_{\ell \in {\mathcal G}(b)}\varphi_j(\lambda_{\ell,b}) \overline{\varphi_k(\lambda_{\ell,b})} \mathring{S}_{r_j s_js_kr_k}(\lambda_{\ell,b})\\
    % \widetilde{f}_{r_jr_k}(\lambda_{l,b}) \widetilde{f}_{s_js_k}(-\lambda_{l,b})\\
 & \ \ \ \ + \frac{4\pi^2}{b} \sum_{\ell \in {\mathcal G}(b)}\varphi_j(\lambda_{\ell,b}) \overline{\varphi_k(-\lambda_{\ell,b})} \mathring{S}_{r_js_jr_ks_k}(\lambda_{\ell,b})+  o_P(1).
  %\sigma^+_{jk} & = 2\pi \Big( \frac{2\pi}{kb} \sum_{l=-[b/2]}^{[b/2]}\varphi_j(\lambda_{l,b}) \overline{\varphi_k(\lambda_{l,b})} \widetilde{f}_{r_jr_k}(\lambda_{l,b}) \widetilde{f}_{s_js_k}(-\lambda_{l,b})\\
 %& \ \ \ \ \ \  \ \ \ \ \ + \frac{2\pi}{kb} \sum_{l=-[b/2]}^{[b/2]}\varphi_j(\lambda_{l,b}) \overline{\varphi_k(-\lambda_{l,b})} \widetilde{f}_{r_js_k}(\lambda_{l,b}) \widetilde{f}_{s_jr_k}(-\lambda_{l,b})\Big).
\end{align*}
and
  \begin{align*}
  c^+_{jk} & = \frac{4\pi^2}{b}  \sum_{\ell \in {\mathcal G}(b)}\varphi_j(\lambda_{\ell,b}) \varphi_k(-\lambda_{\ell,b}) \mathring{S}_{r_js_j s_k r_k}(\lambda_{\ell,b})\\
 & \ \ \ \  + \frac{4\pi^2}{b} \sum_{\ell\in {\mathcal G}(b)} \varphi_j(\lambda_{\ell,b}) \varphi_k(\lambda_{\ell,b}) \mathring{S}_{r_js_jr_ks_k}(\lambda_{\ell,b}) + o_P(1),
\end{align*}
where
%\[ S_{rsuw}(\lambda) = \frac{1}{n-b+1}\sum_{t=1}^{n-b+1}\big( I_{t,rs}(\lambda)-\widetilde{f}_{rs}(\lambda)\big) \big( I_{t,uw}(\lambda) - \widetilde{f}_{uw}(\lambda) \big),\]
\[ \mathring{S}_{rsuw}(\lambda) = \frac{1}{n-b+1}\sum_{t=1}^{n-b+1}\big( I_{t,rs}(\lambda)-\widetilde{f}_{rs}(\lambda)\big) \big( I_{t,uw}(\lambda) - \widetilde{f}_{uw}(\lambda) \big).\]
From Lemma~\ref{le.BootLem1} we then conclude that $ \sigma^+_{jk}\stackrel{P}{\rightarrow} \Sigma_{1;jk} $ and  that $  c^+_{jk}\stackrel{P}{\rightarrow} \Gamma_{1;jk} $.
\hfil $\Box$

\

{\sc Proof of Lemma~\ref{le.BootLem2} $(iii)$:}
We use the notation $ D^*_{rs}(\lambda)=I_{rs}^\ast(\lambda) - \widehat{f}_{rs}(\lambda)$. To show that the $J$-dimensional complex vector
\[ {\bf V}^\ast_n=\Big(\frac{2\pi}{\sqrt{n}} \sum_{l\in{\mathcal G}(n)}\varphi_{j}(\lambda_{l,n})D_{r_js_j}^\ast(\lambda_{l,n}), j=1,2, \ldots, J\Big)^\top, \]
converges weakly to the complex normal variable $ {\mathcal N}_J^c({\bf 0}, {\bf \Sigma}_1, {\bf \Gamma}_1)$, it suffices to show that the $2J$-dimensional real vector
\[ {\bf V}^\ast_{n,T}  = \frac{2\pi}{\sqrt{n}} \sum_{l\in{\mathcal G}(n)} \left(\begin{array}{c}
{\rm Re}\{\varphi_1(\lambda_{l,n})D_{r_1s_1}^\ast(\lambda_{l,n})\}\\
\vdots \\
{\rm Re}\{\varphi_J(\lambda_{l,n})D_{r_Js_J}^\ast(\lambda_{l,n})\}\\
{\rm Im}\{\varphi_1(\lambda_{l,n})D_{r_1s_1}^\ast(\lambda_{l,n})\}\\
\vdots \\
{\rm Im}\{\varphi_J(\lambda_{l,n})D_{r_Js_J}^\ast(\lambda_{l,n})\}\\
\end{array} \right)\]
converges to the $2J$ dimensional,  real normal distribution $ {\mathcal N}_{2J}({\bf 0}, {\bf \Sigma}_V)$, where ${\bf \Sigma}_V$ denotes the limit
of $E^\ast({\bf V}^\ast_{n,T} {\bf V}^{\ast \top}_{n,T})$, which in view of assertion (i) is well defined. Notice that  the covariance matrix $ {\bf \Sigma}_1$ and the relation matrix $ {\bf \Gamma}_1$ together specify $ {\bf \Sigma}_V$ and vice versa. To  simplify the presentation we give only   the proof for the case where the functions $ \varphi_j$ are real-valued.
The case of  complex-valued $\varphi_j$ can be  proved  along  the same lines but with a much more  complicated notation. Write
${\bf V}_{n,T}^\ast = \sum_{l=1}^N  {\bf Z}^\ast_l$,  where  the $ j $-th component of ${\bf Z}^\ast_l$ is given for  $j=1,2, \ldots, J$, by
$$ \frac{2\pi}{\sqrt{n}} \Big({\rm Re}\{\varphi_j(\lambda_{l,n})D_{r_js_j}^\ast(\lambda_{l,n})\} + {\rm Re}\{\varphi_j(-\lambda_{l,n})D_{r_js_j}^\ast(-\lambda_{l,n})\} \Big)$$
and for $ j=J+1, \ldots, 2J$, by
$$   \frac{2\pi}{\sqrt{n}} \Big({\rm Im}\{\varphi_{j-J}(\lambda_{l,n})D_{r_{j-J}s_{j-J}}^\ast(\lambda_{l,n})\} + {\rm Im}\{\varphi_{j-J}(-\lambda_{l,n})D_{r_{j-J}s_{j-J}}^\ast(-\lambda_{l,n})\}\Big). $$
Observe that the random vectors $ {\bf Z}^\ast_l$ are independent.  Thus in view of assertion (i),  to establish the desired weak convergence,  it suffices
to show that  Lyapunov's condition is satisfied, that is, that $ \sum_{l=1}^NE^\ast \|{\bf Z}^\ast_l\|^{2+\delta} \rightarrow 0$ for some $ \delta >0$. Choose $ \delta=2$. Using $ (a+b)^2\leq 2a^2 + 2b^2$
and $\varphi(\lambda) {\rm Re}\{D_{rs}^\ast(\lambda)\}^2 +  \varphi(\lambda) {\rm Im}\{D_{rs}^\ast(\lambda)\}^2= \varphi(\lambda)|D^\ast_{rs}(\lambda)|^2$ we have
\begin{align*}
\sum_{l=1}^NE^\ast \|{\bf Z}^\ast_l\|^{4} & \leq  \frac{(2\pi)^4}{n^2}\sum_{l=1}^NE^\ast\Big( 2\sum_{j=1}^{J}\varphi_{j}^2(\lambda_{l,n})|D^\ast_{r_js_j}(\lambda_{l,n})|^2 \\
& \ \ \ \ \ \ \ \ +  2\sum_{j=1}^{J}\varphi_{j}^2(-\lambda_{l,n})|D^\ast_{r_js_j}(-\lambda_{l,n})|^2 \Big)^2\\
& \leq  \frac{4(2\pi)^4}{n^2} \max_{1\leq j \leq J} \sup_{\lambda\in[-\pi,\pi]} \varphi^2_j(\lambda) \sum_{l=1}^NE^\ast\Big(\sum_{j=1}^J|D^*_{r_js_j}(\lambda_{l,n})|^2 \\
& \ \ \ \ \ \ \  \  + \sum_{j=1}^J|D^*_{r_js_j}(-\lambda_{l,n})|^2\Big)^2\\
& \leq \frac{8(2\pi)^4}{n^2} \max_{1\leq j \leq J} \sup_{\lambda\in[-\pi,\pi]} \varphi^2_j(\lambda) \Big(\sum_{l=1}^N \sum_{j=1}^JE^\ast|D^*_{r_js_j}(\lambda_{l,n})|^4 \\
& \ \ \ \ \ \ \ \  +\sum_{l=1}^N \sum_{j=1}^J E^\ast|D^*_{r_js_j}(-\lambda_{l,n})|^4\Big).
\end{align*}
Now, recall the definition of  $ D^*_{r_js_j}(\lambda)$ and verify  by staightforward calculations that
\begin{align*}
E^\ast |I^*_{rs}(\lambda)-& \widehat{f}_{rs}(\lambda)|^4  =  E^\ast[I^*_{rs}(\lambda)^2I^*_{sr}(\lambda)^2] +  + \widehat{f}_{rs}(\lambda)^2\widehat{f}_{sr}(\lambda)^2 \\
&  - 2\widehat{f}_{rs}(\lambda) E^\ast[I^*_{rs}(\lambda) I^*_{sr}(\lambda)^2] - 2\widehat{f}_{sr}(\lambda)E^\ast[ I^*_{rs}(\lambda)^2 I^*_{sr}(\lambda)]  \\
& + \widehat{f}_{rs}(\lambda)^2 E^\ast[I^*_{sr}(\lambda)^2] + \widehat{f}_{sr}(\lambda)^2 E^\ast[I^*_{rs}(\lambda)^2] \\
& + 4 \widehat{f}_{rs}(\lambda) \widehat{f}_{sr}(\lambda) E^\ast[I^*_{rs}(\lambda) I^*_{sr}(\lambda)] \\
& - 2 \widehat{f}_{rs}(\lambda)\widehat{f}_{sr}(\lambda)^2 E^\ast[I^*_{rs}(\lambda)] -
 2 \widehat{f}_{rs}(\lambda)^2 \widehat{f}_{sr}(\lambda)E^\ast[I^*_{sr}(\lambda)]\,.
\end{align*}
%%%%%%%%%%%%%%%%%%%%%%%%%%%%% Old version:
%\begin{align*}
%E^\ast |I^*_{rs}(\lambda)-& \widehat{f}_{rs}(\lambda)|^4  =  E^\ast[I^*_{rs}(\lambda)I^*_{sr}(\lambda)I^*_{rs}(\lambda))I^*_{sr}(\lambda)] \\
%&  - \widehat{f}_{rs}(\lambda)\big(E^\ast[ I^*_{rs}(\lambda)I^*_{sr}(\lambda)I^*_{rs}(\lambda))] + 2 E^\ast[I^*_{rs}(\lambda)I^*_{sr}(\lambda)I^*_{sr}(\lambda)] \big)\\
%& - \widehat{f}_{sr}(\lambda)E^\ast[ I^*_{rs}(\lambda)I^*_{rs}(\lambda)I^*_{sr}(\lambda)] +2\widehat{f}_{rs}(\lambda)\widehat{f}_{sr}(\lambda) E^\ast[ I^*_{sr}(\lambda)I^*_{rs}(\lambda)]\\
%& + \widehat{f}_{rs}(\lambda)\widehat{f}_{sr}(\lambda) \big(E^\ast[I^*_{rs}(\lambda)I^*_{rs}(\lambda)] + E^\ast[I^*_{sr}(\lambda)I^*_{sr}(\lambda)]\big)\\
%& - 2 \widehat{f}_{sr}(\lambda)\widehat{f}_{rs}(\lambda)\widehat{f}_{sr}(\lambda)E^\ast[I^*_{rs}(\lambda)] -
% 2 \widehat{f}_{rs}(\lambda)\widehat{f}_{rs}(\lambda)\widehat{f}_{sr}(\lambda)E^\ast[I^*_{sr}(\lambda)] \\
% & + \widehat{f}_{rs}(\lambda)\widehat{f}_{sr}(\lambda)\widehat{f}_{rs}(\lambda)\widehat{f}_{sr}(\lambda).
%\end{align*}
Now since $ {\bf I}^\ast(\lambda_{l,n}) $ has  for every $ l=1,2, \ldots, N$ a complex Wishart distribution with parameters $1$ and $ \widehat{\bf f}(\lambda_{l,n}) $,  for short,   ${\bf I}^\ast(\lambda_{l,n}) \sim {\rm Wishart}\big(1,\widehat{\bf f}(\lambda_{l,n})\big)$, we get using Assumption \ref{assu5} and expressions for the moments of the complex Wishart distribution, see Withers and Nadarajah (2012),  that
$ E^\ast |I^*_{rs}(\lambda)- \widehat{f}_{rs}(\lambda)|^4  \stackrel{P}{\rightarrow} E | S_{rs}(\lambda) - f_{rs}(\lambda)|^4 ={\mathcal O}(1)$, for all $ r,s \in \{1,2, \ldots, m\}$, where the random variable $ {\bf S}(\lambda)=(S_{rs}(\lambda))_{r,s=1,2, \ldots, m}$ has the  $ {\rm Wishart}(1,{\bf f}(\lambda))$ distribution.  Therefore,
\[  \sum_{l=1}^NE^\ast \|{\bf Z}^\ast_l\|^{4}  \leq \frac{8(2\pi)^4}{n^2} \max_{1\leq j \leq J} \sup_{\lambda\in[-\pi,\pi]} \varphi^2_j(\lambda)\cdot N\cdot  {\mathcal O}_P(1) = {\mathcal O}_P(n^{-1}) \rightarrow 0.\]
$ \hfill \square $

\

{\sc Proof of Theorem~\ref{th.1}:}
For assertion $(i)$, note that $ E^*({\bf V}_n^\circ)=E^*({\bf V}_n^*) = {\bf 0} $ which implies $ {\rm Cov}^*(({\rm Re}({\bf V}_n^*)^\top,{\rm Im}({\bf V}_n^*)^\top)^\top)= {\bf G}_n^* $. From the definition of $ {\bf C}_n^+ $ one can see that $ {\bf C}_n^+ $, and thus also $ {\bf G}_n^\circ $, are symmetric. It follows from \eqref{eq.Co} that
\bee
{\rm Cov}^*\left[ \begin{pmatrix}
{\rm Re}({\bf V}_n^\circ)\\
{\rm Im}({\bf V}_n^\circ)
\end{pmatrix} \right] = \big({\bf G}_{n}^\circ \big)^{1/2}\big({\bf G}^\ast_n \big)^{-1/2} {\bf G}^\ast_n \big({\bf G}^\ast_n \big)^{-1/2} \big({\bf G}_{n}^\circ \big)^{1/2} = {\bf G}_{n}^\circ \stackrel{P}{\rightarrow} {\bf G}\,,
\eee
due to Lemma~\ref{le.BootLem2}, \eqref{equiv_complexreal} and \eqref{defG}. Again using \eqref{equiv_complexreal} yields assertion $(i)$. Since $ ({\rm Re}({\bf V}_n^*)^\top,{\rm Im}({\bf V}_n^*)^\top)^\top $ is asymptotically normal due to Lemma~\ref{le.BootLem2} $ (iii) $, $ ({\rm Re}({\bf V}_n^\circ)^\top,{\rm Im}({\bf V}_n^\circ)^\top)^\top $ is asymptotically normal with covariance matrix $ {\bf G} $ which gives assertion $ (ii) $. $ \hfill \square $

\

{\sc Proof of Theorem~\ref{th.2}:}
First note that $ \sqrt{n}({\bf M}^*_n -\widehat{\bf M}_n ) $ equals $ {\bf V}^*_n $ from Step I.3. Hence, Lemma~\ref{le.BootLem2} $ (iii) $ together with \eqref{equiv_complexreal} yields
\be
\sqrt{n}\left[\begin{pmatrix}
{\rm Re}({\bf M}^*_n)\\
{\rm Im}({\bf M}^*_n)
\end{pmatrix} -\begin{pmatrix}
{\rm Re}(\widehat{\bf M}_n)\\
{\rm Im}(\widehat{\bf M}_n)
\end{pmatrix} \right] \stackrel{d}{\longrightarrow} \mathcal N_{2J}({\bf 0},{\bf G_1}) \label{Th39_h1}
\ee
in $ P $-probability, where $ {\bf G_1} $ is defined analogous to $ {\bf G} $ in \eqref{defG} but with $ {\bf \Sigma} $ replaced by $ {\bf \Sigma_1} $ and $ {\bf \Gamma} $ replaced by $ {\bf \Gamma_1} $. We next apply the delta method to \eqref{Th39_h1}. Invoking the mean value theorem for all $ 2L $ component functions of $ \tilde g $ leads to
\begin{align}
\begin{pmatrix}
{\rm Re}({\bf W}^*_n)\\
{\rm Im}({\bf W}^*_n)
\end{pmatrix} = \big( \nabla \tilde g_i(\xi_i) \big)_{i=1,\ldots,2L} \times \sqrt{n}\left[\begin{pmatrix}
{\rm Re}({\bf M}^*_n)\\
{\rm Im}({\bf M}^*_n)
\end{pmatrix} -\begin{pmatrix}
{\rm Re}(\widehat{\bf M}_n)\\
{\rm Im}(\widehat{\bf M}_n)
\end{pmatrix} \right]\,, \label{Th39_h2}
\end{align}
where $ \nabla \tilde g_i(\cdot) $ denotes the gradient (as a row vector) of the $ i $-th component function of $ \tilde g $, and $ \xi_i \in \R^{2J} $ denotes a vector on the line segment between $ ({\rm Re}({\bf M}^*_n)^\top,{\rm Im}({\bf M}^*_n)^\top)^\top $ and $ ({\rm Re}(\widehat{\bf M}_n)^\top,{\rm Im}(\widehat{\bf M}_n)^\top)^\top $. We now have from \eqref{Th39_h1} that $ \| \xi_i- ({\rm Re}(\widehat{\bf M}_n)^\top,{\rm Im}(\widehat{\bf M}_n)^\top)^\top \| \rightarrow 0 $ in $ P^* $-probability for all $ i \in \{1,\ldots,2L\} $, and Assumption \ref{assu5} together with convergence of Riemann sums yields $ \| \widehat{\bf M}_n - {\bf M} \| \stackrel{P}{\rightarrow} 0 $. Therefore \eqref{Th39_h2} and \eqref{Th39_h1} imply
\begin{align}
\begin{pmatrix}
{\rm Re}({\bf W}^*_n)\\
{\rm Im}({\bf W}^*_n)
\end{pmatrix} \stackrel{d}{\longrightarrow} {\bf J}_{\tilde g}(( {\rm Re}({\bf M})^\top,{\rm Im}({\bf M})^\top)^\top) \, \mathcal{N}_{2J}({\bf 0},{\bf G_1}) \label{Th39_h3}
\end{align}
in $ P $-probability. We only need the asymptotic normality from this statement. Note that $ \widetilde {\bf G}_n^* $ is the covariance matrix of the last left-hand side and therefore
\begin{align}
\big( \widetilde {\bf G}_n^*\big)^{-1/2} \begin{pmatrix}
{\rm Re}({\bf W}^*_n)\\
{\rm Im}({\bf W}^*_n)
\end{pmatrix} \stackrel{d}{\longrightarrow} \mathcal{N}_{2J}({\bf 0},\UMa_{2J})
\end{align}
in probability. From the proof of Theorem~\ref{th.1} we have $ {\bf G}_{n}^\circ \stackrel{P}{\rightarrow} {\bf G} $. Also, the above considerations together with continuity of $ {\bf J}_{\tilde g} $ yield $ {\bf J}_{\tilde g}(({\rm Re}(\widehat{\bf M}_n)^\top,{\rm Im}(\widehat{\bf M}_n)^\top)^\top) \stackrel{P}{\rightarrow} {\bf J}_{\tilde g}(({\rm Re}({\bf M})^\top,{\rm Im}({\bf M})^\top)^\top) $. Thus,
\begin{align*}
\widetilde{\bf G}^\circ_n \stackrel{P}{\longrightarrow} {\bf J}_{\tilde g}\begin{pmatrix}{\rm Re}({\bf M})\\ {\rm Im}({\bf M})\end{pmatrix} \times {\bf G} \times {\bf J}_{\tilde g}\begin{pmatrix}{\rm Re}({\bf M})\\ {\rm Im}({\bf M})\end{pmatrix}^\top\,.
\end{align*}
Therefore we have
\begin{align}
\begin{pmatrix}
{\rm Re}({\bf W}^\circ_n)\\
{\rm Im}({\bf W}^\circ_n)
\end{pmatrix} \stackrel{d}{\longrightarrow} {\bf J}_{\tilde g}(( {\rm Re}({\bf M})^\top,{\rm Im}({\bf M})^\top)^\top) \, \mathcal{N}_{2J}({\bf 0},{\bf G})
\end{align}
in probability. Applying the continuous mapping theorem with $ h(x_1,x_2):=x_1+i\cdot x_2 $, for $ x_1,x_2 \in \R^{L} $, it follows
\begin{align}
{\bf W}^\circ_n \stackrel{d}{\longrightarrow} {\bf W} \sim \mathcal{N}^c_{L}({\bf 0},{\bf \Sigma_R},{\bf \Gamma_R})
\end{align}
in probability, for the matrices $ {\bf \Sigma_R},{\bf \Gamma_R} $ stated in \eqref{CLT_smoothfun_g} (the exact form of which does not need to be specified for the desired assertion to hold). $ \hfill \square $

%\clearpage

%%%%%%%%%%%%%%%%%%%%%%%%%%%% SUPPLEMENT %%%%%%%%%%%%%%%%%%%%%%%%%%%%%%%%%%%%%%%%%%

\clearpage

\begin{center}
\textbf{A FREQUENCY DOMAIN BOOTSTRAP FOR GENERAL\\MULTIVARIATE STATIONARY PROCESSES}

\

\textbf{-- SUPPLEMENTARY MATERIAL --}
\end{center}

\

{\sc Proof of Lemma \ref{lemmacovper}:} The proof generalizes calculations from Rosenblatt (1985) and Krogstad (1982) regarding the covariance structure of univariate periodogram ordinates. The assertion of Lemma \ref{lemmacovper} remains true if one switches from Fourier frequencies $ \lambda_{j,n}, \lambda_{k,n} $ to two fixed frequencies $ \lambda_1,\lambda_2\in [0,\pi] $; we will state at the end of this proof which arguments have to be adapted in this situation.\\
A direct calculation yields the decomposition of the covariance into three major components:
\bee
& & {\rm Cov}(I_{rs}(\lambda_{j,n}),I_{vw}(\lambda_{k,n})) \\
&=& \frac{1}{4\pi^2 n^2} \sum_{t_1,t_2,t_3,t_4=1}^{n} {\rm cum}(X_r(t_1),X_s(t_2),X_v(t_3),X_w(t_4)) \, e^{-i(t_1-t_2)\lambda_{j,n}+i(t_3-t_4)\lambda_{k,n}} \\
& & + \frac{1}{4\pi^2 n^2} \sum_{t_1,t_2,t_3,t_4=1}^{n} \gamma_{rv}(t_1-t_3) \gamma_{sw}(t_2-t_4) \, e^{-i(t_1-t_2)\lambda_{j,n}+i(t_3-t_4)\lambda_{k,n}} \\
& & + \frac{1}{4\pi^2 n^2} \sum_{t_1,t_2,t_3,t_4=1}^{n} \gamma_{rw}(t_1-t_4) \gamma_{sv}(t_2-t_3) \, e^{-i(t_1-t_2)\lambda_{j,n}+i(t_3-t_4)\lambda_{k,n}} \\
&=:& S_1+S_2+S_3\,.
\eee
For $ S_1 $ we obtain with an index shift for three summands
\bee
\frac{n}{2\pi}S_1 &=& \frac{1}{(2\pi)^3 n} \sum_{t_1,t_2,t_3,t_4=1}^{n} c_{rsvw}(t_1-t_4,t_2-t_4,t_3-t_4) \, e^{-i(t_1-t_2)\lambda_{j,n}+i(t_3-t_4)\lambda_{k,n}} \\
&=& \frac{1}{(2\pi)^3 n} \sum_{t_4=1}^{n} \sum_{t_1,t_2,t_3=1-t_4}^{n-t_4} c_{rsvw}(t_1,t_2,t_3) \, e^{-i(t_1-t_2)\lambda_{j,n}+it_3\lambda_{k,n}}\,.
\eee
By merging like summands the last expression can be seen to be equal to
\begin{align}
\frac{1}{(2\pi)^3 n} \sum_{h_1,h_2,h_3=-(n-1)}^{n-1} q_n(h_1,h_2,h_3)\, c_{rsvw}(h_1,h_2,h_3) \, e^{-i(h_1-h_2)\lambda_{j,n}+ih_3\lambda_{k,n}}\,, \label{prooflem1}
\end{align}
where $ q_n(h_1,h_2,h_3) $ counts the number of times the respective summand appears in $ (n/2\pi)S_1 $. It holds
\bee
q_n(h_1,h_2,h_3) = \Big( n - \max\{ |h_1|,|h_2|,|h_3|,|h_1-h_2|,|h_1-h_3|,|h_2-h_3| \} \Big)_+\,.
\eee
We can replace the $ q_n(\cdot) $ term in \eqref{prooflem1} by $ n $ since the resulting remainder term vanishes with rate $ \mathcal{O}(n^{-1}) $ as the following bound shows:
\bee
& & \left|\frac{1}{(2\pi)^3 n} \sum_{h_1,h_2,h_3=-(n-1)}^{n-1} \big(q_n(h_1,h_2,h_3)-n\big) c_{rsvw}(h_1,h_2,h_3) \, e^{-i\ldots} \right| \\
&\leq & \frac{1}{(2\pi)^3 n} \sum_{h_1,h_2,h_3=-\infty}^{\infty} 2 \, \max\{|h_1|,|h_2|,|h_3|\} \, |c_{rsvw}(h_1,h_2,h_3)| \\
&\leq & \frac{1}{(2\pi)^3 n} \sum_{h_1,h_2,h_3=-\infty}^{\infty} 2 \, (1+|h_1|+|h_2|+|h_3|) \, |c_{rsvw}(h_1,h_2,h_3)|\,,
\eee
which is of order $ \mathcal{O}(n^{-1}) $. Now, after replacing $ q_n(\cdot) $ with $ n $ in \eqref{prooflem1}, the difference of the remaining term and $ f_{rsvw}(\lambda_{j,n},-\lambda_{j,n},-\lambda_{k,n}) $ can be bounded by
\bee
\frac{1}{(2\pi)^3} \sum_{h_1,h_2,h_3=-\infty}^{\infty} \mathds{1}_{\{\max\{|h_1|,|h_2|,|h_3|\}\geq n \}} |c_{rsvw}(h_1,h_2,h_3)|\,,
\eee
which also vanishes with rate $ \mathcal{O}(n^{-1}) $ under the imposed summability assumption on $ |c_{rsvw}(h_1,h_2,h_3)| $. This yields the desired assertion for $ S_1 $.\\
To handle $ S_2 $ and $ S_3 $ we have to introduce some notation and preliminaries. We define the functions $ g_n(\lambda):= \sum_{k=1}^{n} e^{ik\lambda} $. For all $ \lambda \not\in 2\pi \mathbb{Z} $, we can use the geometric sum formula to express $ g_n $ as
\bee
g_n(\lambda)= e^{i\lambda} \, \frac{e^{in\lambda}-1}{e^{i\lambda}-1} = e^{i(n+1)\lambda/2} \, \frac{\sin(n\lambda/2)}{\sin(\lambda/2)}\,.
\eee
Note that for all $ \lambda \in \R $ we have $ g_n(\lambda) \, g_n(-\lambda)=2\pi n\, F_n(\lambda) $, where $ F_n $ is the (continuously extended) Fej$\acute{{\rm e}}$r kernel:
\bee
F_n(\lambda):= \begin{cases}
\frac{1}{2\pi n} \frac{\sin^2(n\lambda/2)}{\sin^2(\lambda/2)}, & \lambda \not\in 2\pi \Z\\
\frac{n}{2\pi}, & \lambda \in 2\pi \Z
\end{cases}\,.
\eee
Integrals over these functions can be handled in an elegant way by determining their Fourier coefficients. For arbitrary $ \lambda_1,\lambda_2 $ define
\bee
z_{n,\lambda_1,\lambda_2}(\alpha):= g_n(\alpha-\lambda_1) \, g_n(-(\alpha-\lambda_2))\,.
\eee
The $ \ell $-th Fourier coefficient of this function is given by
\be
\widehat z_{n,\lambda_1,\lambda_2}[\ell] &=& \frac{1}{2\pi} \int_{-\pi}^{\pi} g_n(\alpha-\lambda_1)\, g_n(-\alpha+\lambda_2)\, e^{-i\ell \alpha}\, d\alpha \nonumber\\
&=& \frac{1}{2\pi} \sum_{p,q=1}^{n} e^{-ip\lambda_1} e^{iq\lambda_2} \int_{-\pi}^{\pi} e^{i(p-q-\ell)\alpha}\, d\alpha \nonumber\\
&=& \sum_{p,q=1}^{n} e^{-ip\lambda_1} e^{iq\lambda_2} \mathds{1}_{\{ p-q=\ell \}}\,. \label{prooflem2}
\ee
In particular it holds $ \widehat z_{n,\lambda_1,\lambda_2}[\ell]=0 $ for all $ |\ell|\geq n $.\\
With this notation, and using the fact that $ \gamma_{rs}(h)=\int_{-\pi}^{\pi} e^{ih\alpha} f_{rs}(\alpha) \, d\alpha $, we can write
\begin{align}
S_2 = \frac{1}{4\pi^2 n^2} \int_{-\pi}^{\pi} z_{n,\lambda_{j,n},\lambda_{k,n}}(\alpha) \, f_{rv}(\alpha)\, d\alpha \cdot \int_{-\pi}^{\pi} z_{n,-\lambda_{j,n},-\lambda_{k,n}}(\alpha) \, f_{sw}(\alpha)\, d\alpha\,. \label{prooflem3}
\end{align}
Consider first the case with $ j=k $, that is, with $ \lambda_{j,n}=\lambda_{k,n} $. Then the last expression simplifies to
\bee
S_2 = \int_{-\pi}^{\pi} F_n(\alpha-\lambda_{j,n}) \, f_{rv}(\alpha)\, d\alpha \cdot \int_{-\pi}^{\pi} F_n(\alpha+\lambda_{j,n}) \, f_{sw}(\alpha)\, d\alpha\,.
\eee
Since $ F_n(\cdot - \lambda_{j,n})=z_{n,\lambda_{j,n},\lambda_{j,n}}(\cdot)/(2\pi n) $, its $ \ell $-th Fourier coefficient can be obtained from \eqref{prooflem2}, it is
\bee
\widehat F_n(\cdot - \lambda_{j,n})[\ell] = \frac{(n-|\ell|)e^{-i\ell \lambda_{j,n}}}{2\pi n} \mathds{1}_{\{|\ell|<n \}}\,.
\eee
The Fourier coefficients of $ f_{rv} $ are given by
\bee
\widehat f_{rv}[\ell] = \frac{1}{2\pi} \int_{-\pi}^{\pi} f_{rv}(\alpha)\, e^{i(-\ell)\alpha} \, d\alpha = \frac{\gamma_{rv}(-\ell)}{2\pi}\,.
\eee
Then we can apply Parseval's Theorem to calculate the first integral in $ S_2 $ as
\bee
\int_{-\pi}^{\pi} F_n(\alpha-\lambda_{j,n}) \, f_{rv}(\alpha)\, d\alpha &=& 2\pi \sum_{\ell=-\infty}^{\infty} \widehat f_{rv}[\ell] \, \overline{\widehat F_n(\cdot - \lambda_{j,n})[\ell]} \\
&=& \frac{1}{2\pi} \sum_{\ell=-(n-1)}^{n-1} \gamma_{rv}(\ell) \, \Big( 1-\frac{|\ell|}{n} \Big) \, e^{-i\ell \lambda_{j,n}}\,.
\eee
Of course, the last result could have been obtained by direct calculations for the expression $ S_2 $ as well. The calculation via Fourier coefficients presented here serves as the basis for the upcoming other cases to consider. In those cases the proposed way via Fourier coefficients seems much more elegant and shorter than a direct calculation.\\
The difference between the right-hand side of the last equation and $ f_{rv}(\lambda_{j,n}) $ can be shown to be of the order $ \mathcal{O}(n^{-1}) $ with standard arguments, using $ \sum_{h \in \Z} (1+|h|)|\gamma_{rv}(h)|<\infty $. The second integral in $ S_2 $ behaves as $ f_{sw}(-\lambda_{j,n}) $ by analogous arguments. Together we have
\bee
\big| S_2 - f_{rv}(\lambda_{j,n})\cdot f_{sw}(-\lambda_{j,n}) \big| = \mathcal{O}(n^{-1})
\eee
for the case $ \lambda_{j,n}=\lambda_{k,n} $.\\
Next we consider the case with $ j \neq k $, that is, with $ \lambda_{j,n}\neq \lambda_{k,n} $ different Fourier frequencies within $ [0,\pi] $. In this case $ \lambda_{k,n}-\lambda_{j,n} = 2\pi(k-j)/n $ is a Fourier frequency with $ \lambda_{k,n}-\lambda_{j,n} \not\in 2\pi \Z $, and thus $ \sum_{q=1}^{n} e^{iq(\lambda_{k,n}-\lambda_{j,n})}=0 $. We can use this fact together with \eqref{prooflem2} to obtain a bound for the Fourier coefficients of $ z_{n,\lambda_{j,n},\lambda_{k,n}} $ for all $ |\ell|< n $ (all other coefficients are zero anyway):
\begin{align}
\Big|\widehat z_{n,\lambda_{j,n},\lambda_{k,n}}[\ell] \Big| &= \left| \sum_{q=1}^{n-|\ell|} e^{iq(\lambda_{k,n}-\lambda_{j,n})} \right| = \left| - \sum_{q=n-|\ell|+1}^{n} e^{iq(\lambda_{k,n}-\lambda_{j,n})} \right| \leq |\ell| \,. \label{prooflem6}
\end{align}
With this bound the first integral in \eqref{prooflem3} can be bounded via Parseval's Theorem by
\bee
\left| \int_{-\pi}^{\pi} z_{n,\lambda_{j,n},\lambda_{k,n}}(\alpha) \, \overline{f_{vr}(\alpha)}\, d\alpha \right| &=& \left| 2\pi \sum_{\ell=-\infty}^{\infty} \widehat z_{n,\lambda_{j,n},\lambda_{k,n}}[\ell] \, \overline{\widehat f_{vr}[\ell]} \right| \\
&\leq & \sum_{\ell=-\infty}^{\infty} (1+|\ell|) \, |\gamma_{rv}(\ell)| < \infty\,.
\eee
The second integral in \eqref{prooflem3} can be bounded analogously. Therefore, $ S_2 $ vanishes with rate $ \mathcal{O}(n^{-2}) $ in this case.\\
The $ S_3 $ term can be treated similar -- but not completely analogous -- to the $ S_2 $ term. It holds
\begin{align}
S_3 = \frac{1}{4\pi^2 n^2} \int_{-\pi}^{\pi} z_{n,\lambda_{j,n},-\lambda_{k,n}}(\alpha) \, f_{rw}(\alpha)\, d\alpha \cdot \int_{-\pi}^{\pi} z_{n,-\lambda_{j,n},\lambda_{k,n}}(\alpha) \, f_{sv}(\alpha)\, d\alpha\,. \label{prooflem4}
\end{align}
In the case $ \lambda_{j,n}=\lambda_{k,n}=0 $, one can proceed anologously to the $ S_2 $ case and see that $ S_3 $ simplifies to
\bee
S_3 = \int_{-\pi}^{\pi} F_n(\alpha) \, f_{rw}(\alpha)\, d\alpha \cdot \int_{-\pi}^{\pi} F_n(\alpha) \, f_{sv}(\alpha)\, d\alpha\,,
\eee
for which we have
\bee
\big| S_3 - f_{rw}(0)\cdot f_{sv}(0) \big| = \mathcal{O}(n^{-1})\,.
\eee
Now let $ \lambda_{j,n}=\lambda_{k,n}=\pi $. This case is slightly more subtle because both integrals in \eqref{prooflem4} have to be treated together to see that certain terms cancel out. We then get
\begin{align}
S_3 = & \int_{-\pi}^{\pi} \frac{1}{2\pi n} \frac{\sin \left(-\frac{n\pi}{2}+\frac{n\alpha}{2}\right) \, \sin \left(-\frac{n\pi}{2}-\frac{n\alpha}{2}\right) }{\sin \left( -\frac{\pi}{2}+\frac{\alpha}{2}\right) \, \sin \left( -\frac{\pi}{2}-\frac{\alpha}{2} \right)} \, f_{rw}(\alpha)\, d\alpha \label{prooflem5}\\
& \quad \times \int_{-\pi}^{\pi} \frac{1}{2\pi n} \frac{\sin \left(-\frac{n\pi}{2}+\frac{n\alpha}{2}\right) \, \sin \left(-\frac{n\pi}{2}-\frac{n\alpha}{2}\right) }{\sin \left( -\frac{\pi}{2}+\frac{\alpha}{2}\right) \, \sin \left( -\frac{\pi}{2}-\frac{\alpha}{2} \right)} \, f_{sv}(\alpha)\, d\alpha \nonumber\,.
\end{align}
Observe that the function $ \sin (-n\pi/2+ \cdot) $ is even if $ n $ is odd and odd if $ n $ is even. Therefore it holds
\bee
\sin \left(-\frac{n\pi}{2}+\frac{n\alpha}{2}\right) \, \sin \left(-\frac{n\pi}{2}-\frac{n\alpha}{2}\right) = \sin^2 \left(\frac{n(\alpha-\pi)}{2}\right) \, s(n)\,,
\eee
where $ s(n)=-1 $ if $ n $ is even and $ s(n)=1 $ if $ n $ is odd. Thus \eqref{prooflem5} simplifies to
\begin{align}
S_3 &= \int_{-\pi}^{\pi} \frac{s(n)}{2\pi n} \frac{\sin^2 \left(\frac{n(\alpha-\pi)}{2}\right) }{\sin^2 \left(\frac{\alpha-\pi}{2}\right) } \, f_{rw}(\alpha)\, d\alpha \cdot \int_{-\pi}^{\pi} \frac{s(n)}{2\pi n} \frac{\sin^2 \left(\frac{n(\alpha-\pi)}{2}\right) }{\sin^2 \left(\frac{\alpha-\pi}{2}\right) } \, f_{sv}(\alpha)\, d\alpha \nonumber\\
&= \int_{-\pi}^{\pi} F_n(\alpha-\pi) \, f_{rw}(\alpha)\, d\alpha \cdot \int_{-\pi}^{\pi} F_n(\alpha-\pi) \, f_{sv}(\alpha)\, d\alpha \nonumber\,,
\end{align}
which implies
\bee
\big| S_3 - f_{rw}(\pi)\cdot f_{sv}(\pi) \big| = \mathcal{O}(n^{-1})\,,
\eee
for the case $ \lambda_{j,n}=\lambda_{k,n}=\pi $.\\
For all other cases, that is, for $ \lambda_{j,n}=\lambda_{k,n} \not\in \{0,\pi\} $ or $ \lambda_{j,n} \neq \lambda_{k,n} $ we have $ \pm (\lambda_{j,n}+\lambda_{k,n}) \not\in 2\pi \Z $ and we get
\bee
\big|\widehat z_{n,\lambda_{j,n},-\lambda_{k,n}}[\ell] \big| \leq |\ell|  \quad \textrm{and} \quad \big|\widehat z_{n,-\lambda_{j,n},\lambda_{k,n}}[\ell] \big| \leq |\ell|\,,
\eee
which implies $ |S_3|=\mathcal{O}(n^{-2}) $, analog.ous to the calculations for $ S_2 $ before. Moreover, an inspection of this proof shows that in all cases the $ \mathcal{O}(\cdot) $ bounds are uniform over all frequencies. Also, if one is interested in $ {\rm Cov}(I_{rs}(\lambda_1),I_{vw}(\lambda_2)) $ for two fixed frequencies $ \lambda_1,\lambda_2\in [0,\pi] $ instead of Fourier frequencies $ \lambda_{j,n}, \lambda_{k,n} $, one can in large parts use the same arguments presented here. The only argument that changes is \eqref{prooflem6}. For fixed frequencies $ \lambda_1 \neq \lambda_2 $ one gets instead of \eqref{prooflem6}
\begin{align}
\Big|\widehat z_{n,\lambda_1,\lambda_2}[\ell] \Big| &= \left| \sum_{q=1}^{n-|\ell|} e^{iq(\lambda_2-\lambda_1)} \right| = \left| e^{i(\lambda_2-\lambda_1)\frac{n-|\ell|+1}{2}} \frac{\sin \left(\frac{(n-|\ell|)(\lambda_2-\lambda_1)}{2}\right)}{ \sin \left(\frac{\lambda_2-\lambda_1}{2}\right) } \right| \leq C_{\lambda_1,\lambda_2} \,,
\end{align}
where the finite constant $ C_{\lambda_1,\lambda_2} $ depends only on $ \lambda_1,\lambda_2 $ but not on $ n $. With this bound one can proceed as in the case of Fourier frequencies. $ \hfill \square $

\

{\sc Proof of Lemma~\ref{le.BootLem1} $(i)$:}
For the variance of $ \widetilde{f}_{rs}(\lambda_{\ell,b})$ we have
\begin{align*}
\textrm{Var}(\widetilde{f}_{rs}(\lambda_{\ell,b})) & = \frac{1}{(n-b+1)^2} \sum_{t_1,t_2=1}^{n-b+1} \textrm{Cov}(I_{t_1;rs}(\lambda_{\ell,b}), I_{t_2;rs}(\lambda_{\ell,b}))\\
%& \leq \frac{1}{(n-b+1)^2}\frac{1}{4\pi^2b^2}\sum_{t_1,t_2=1}^{n-b+1}\sum_{g_1,g_2,g_3,g_4=1}^b\\
% \Big\{\big|Cov(X_r(t_1+g_1-1),& X_r(t_2+ g_3-1))\big| \big|Cov(X_s(t_1+g_2-1),X_s(t_2+g_4-1)\big|\\
% + \big|Cov(X_r(t_1+g_1-1),& X_r(t_2+ g_4-1))\big| \big|Cov(X_s(t_1+g_2-1),X_r(t_2+g_3-1)\big|\\
% + \big|cum(X_r(t_1+g_1-1),& X_s(t_1+ g_2-1)), X_r(t_2+g_3-1),X_s(t_2+g_4-1)\big|\Big\}\\
 & \leq \frac{1}{(n-b+1)^2}\frac{1}{4\pi^2b^2}\sum_{t_1,t_2=1}^{n-b+1}\sum_{g_1,g_2,g_3,g_4=1}^b\\
  & \Big\{\big|\gamma_{rr}((t_1-t_2) +(g_1-g_3))\big|  \big|\gamma_{ss}((t_1-t_2) +(g_2-g_4))\big|  \\
 & +\big|\gamma_{rs}((t_1-t_2) +(g_1-g_4))\big|\big|\gamma_{sr}((t_1-t_2) +(g_2-g_3))\big|  \\
  & + \big| c_{rsrs}(t_1-t_2+g_1-g_4,t_1-t_2+g_2-g_4,g_3-g_4) \big|\Big\}.
\end{align*}
For the first term on the right hand side of the last expression, we get by first summing over $g_4$ and then over $t_1$ that this term is ${\mathcal O}(b/(n-b+1))$.
The ${\mathcal O}(b/(n-b+1))$  bound  is obtained for the second term by the same arguments  while for the last term we get by summing first over $g_1$, then over $g_3$ and then over $t_1$ that this term is $ {\mathcal O}(1/(n-b+1))$.  Now, since $ E(\widetilde{f}_{rs}(\lambda_{\ell,b}) )= E(I_{1;rs}(\lambda_{\ell,b}))$, we get that
\begin{align*}
E\Big( \sum_{\ell\in {\mathcal G}(b)}\big|\widetilde{f}_{rs}(\lambda_{\ell,b})  - &E(I_{1;rs}(\lambda_{\ell,b}))\big| \Big)
\leq \sum_{\ell\in{\mathcal G}(b)}\sqrt{\textrm{Var}(\widetilde{f}_{r,s}(\lambda_{\ell,b}))} \\
%& = b{\mathcal O}(\sqrt{b/(n-b+1)}) \\
& = {\mathcal O}(\sqrt{b^3/(n-b+1)})\,,
\end{align*}
which completes the proof. $ \hfill \square $

\

{\sc Proof of Lemma~\ref{le.BootLem1} $(ii)$:}
Notice that
\begin{align*}
&\frac{\displaystyle 1}{\displaystyle n-b+1}\sum_{t=1}^{n-b+1} \Big( I_{t;r_js_j}(\lambda_{\ell_1,b})I_{t;r_ks_k}(\lambda_{\ell_2,b})   - E\big(I_{t;r_js_j}(\lambda_{\ell_1,b})I_{t;r_ks_k}(\lambda_{\ell_2,b}) \big)\Big)  \\
 = &\; E\big(I_{1; r_ks_k}(\lambda_{\ell_2,b}) \big)  \frac{\displaystyle 1}{\displaystyle n-b+1}\sum_{t=1}^{n-b+1}   \Big( I_{t;r_js_j}(\lambda_{\ell_1,b})  - E\big(I_{t;r_js_j}(\lambda_{\ell_1,b}) \big)\Big)\\
 & +E\big(I_{1; r_js_j}(\lambda_{\ell_1,b}) \big) \frac{\displaystyle 1}{\displaystyle n-b+1}\sum_{t=1}^{n-b+1}   \Big( I_{t;r_ks_k}(\lambda_{\ell_2,b})  - E\big(I_{t;r_ks_k}(\lambda_{\ell_2,b}) \big)\Big)\\
 & + \frac{\displaystyle 1}{\displaystyle n-b+1}\sum_{t=1}^{n-b+1} \\
   & \; \Big\{  \big( I_{t;r_js_j}(\lambda_{\ell_1,b}) - E\big(I_{t;r_js_j}(\lambda_{\ell_1,b}) \big)\big) \big( I_{t;r_ks_k}(\lambda_{\ell_2,b})  - E\big(I_{t;r_ks_k}(\lambda_{\ell_2,b}) \big)\big)\\
& - \; E\Big(\big( I_{t;r_js_j}(\lambda_{\ell_1,b}) - E\big(I_{t;r_js_j}(\lambda_{\ell_1,b}) \big)\big)
 \big( I_{t;r_ks_k}(\lambda_{\ell_2,b})  - E\big(I_{t;r_ks_k}(\lambda_{\ell_2,b}) \big)\big)\Big)\Big\}\\
 =: & \;  T_{1,n} (\ell_1, \ell_2)+ T_{2,n}(\ell_1,\ell_2)  + T_{3,n}(\ell_1,\ell_2)
\end{align*}
with an obvious notation for $ T_{j,n}(\ell_1,\ell_2)$, $ j=1,2,3$. Using $ E\big(I_{1; rs}(\lambda_{\ell,b}) \big)= f_{rs}(\lambda_{\ell,b}) + {\mathcal O}(1/b)$ and
assertion $(i)$ of the lemma, we get
\begin{align*}
\sum_{\ell_1,\ell_2\in{\mathcal G}(b)}& | T_{1,n}(\ell_1,\ell_2)|   \leq \sum_{m_ 2\in{\mathcal G}(b)} \big|  E\big(I_{1;r_ks_k}(\lambda_{\ell_2,b}) \big)\big|\\
& \ \ \ \times \sum_{\ell_1\in{\mathcal G}(b)} \Big| \frac{\displaystyle 1}{\displaystyle n-b+1}\sum_{t=1}^{n-b+1}   \Big( I_{t;r_js_j}(\lambda_{\ell_1,b})  - E\big(I_{t;r_js_j}(\lambda_{\ell_1,b}) \big)\Big)\Big| \\
& = {\mathcal O}_{P}(\sqrt{b^5/(n-b+1)}).
\end{align*}
By analogous arguments, the same bound is obtained for the term $ T_{2,n}(\ell_1,\ell_2)$. For the term  $ T_{3,n}(\ell_1,\ell_2)$ observe that  $ E(T_{3,n}(\ell_1,\ell_2))=0$ and
that $ \sum_{\ell_1,\ell_2\in{\mathcal G}(b)}E |T_{3,n}(\ell_1,\ell_2)| \leq \sum_{\ell_1,\ell_2\in {\mathcal G}(b)} \sqrt{\textrm{Var}(T_{3,n}(\ell_1,\ell_2))}$.
Using the  notation $ I^c_{t;rs}(\lambda)=
I_{t;rs}(\lambda)- E(I_{t;rs}(\lambda))$ we get  for the variance
\begin{align*}
\textrm{Var}( &T_{3,n}(\ell_1,\ell_2)) =  \frac{\displaystyle 1}{\displaystyle (n-b+1)^2}\sum_{t_1,t_2=1}^{n-b+1} \\
& \Big\{ E\big(I^c_{t_1;r_js_j}(\lambda_{\ell_1,b}) \overline{I}^c_{t_2;r_js_j}(\lambda_{\ell_1,b})\big)E\big(I^c_{t_1;r_ks_k}(\lambda_{\ell_2,b})\overline{I}^c_{t_2;r_ks_k}(\lambda_{\ell_2,b})\big)\\
 & +  E\big(I^c_{t_1;r_js_j}(\lambda_{\ell_1,b}) \overline{I}^c_{t_2;r_ks_k}(\lambda_{\ell_2,b})\big)E\big(\overline{I}^c_{t_2;r_js_j}(\lambda_{\ell_1,b}) I^c_{t_1;r_ks_k}(\lambda_{\ell_2,b})\big)\\
 & + \textrm{cum}\big( I^c_{t_1;r_js_j}(\lambda_{\ell_1,b}) , I^c_{t_1;r_ks_k}(\lambda_{\ell_2,b}), \overline{I}^c_{t_2;r_js_j}(\lambda_{\ell_1,b}),\overline{I}^c_{t_2;r_ks_k}(\lambda_{\ell_2,b})\big)\Big\}\\
 =: &\; V_{1,n}(\ell_1,\ell_2) + V_{2,n}(\ell_1,\ell_2) + V_{3,n}(\ell_1,\ell_2),
\end{align*}
with an obvious notation for $ V_{1,n}(\ell_1,\ell_2)$, $j=1,2,3$. We show that  each one of the terms   $ V_{j,n}(\ell_1,\ell_2)$  is of the order $ {\mathcal O}(b/(n-b+1))$. This  implies that
\[ \sum_{\ell_1,\ell_2\in {\mathcal G}(b)}  \sqrt{V_{1,n}(\ell_1,\ell_2) + V_{2,n}(\ell_1,\ell_2) + V_{3,n}(\ell_1,\ell_2)} = {\mathcal O}(\sqrt{b^5/(n-b+1)}).\]
The terms $ V_{1,n}(\ell_1,\ell_2)$ and $V_{2,n}(\ell_1,\ell_2) $ can be treated similarly, so we only consider $ V_{1,n}(\ell_1,\ell_2)$. For this term we have using
\begin{align*}
\Big|E(I^c_{t_1;r_js_j}&(\lambda_{\ell_1,b}) \overline{I}^c_{t_2;r_js_j}(\lambda_{\ell_1,b})\big) \Big| \leq \frac{1}{4\pi^2 b^2}
\sum_{g_1,g_2,g_3,g_4=1}^b \Big\{ \\
& |\gamma_{r_jr_j}((t_1-t_2)+(g_1-g_3)) \gamma_{s_js_j}((t_1-t_2)+(g_2-g_4))| \\
+ & |\gamma_{r_js_j}((t_1-t_2)+(g_1-g_4))\gamma_{s_jr_j}((t_1-t_2)+(g_2-g_3))| \\
+ & \big| c_{r_js_jr_js_j}( t_1-t_2+g_1-g_4,t_1-t_2+g_2-g_4,g_3-g_4 ) \big| \Big\}
\end{align*}
that
\begin{align}
\big|V_{1,n}(\ell_1,\ell_2)\big| \leq  \frac{1}{16 \pi^4 b^4}\frac{1}{(n-b+1)^2} \sum_{t_1,t_2=1}^{n-b+1} C_j(t_1,t_2) \, C_k(t_1,t_2)\,, \label{lem_h1}
\end{align}
where
\begin{align*}
C_j(t_1,t_2):= &\sum_{g_1,g_2,g_3,g_4=1}^b\Big\{ |\gamma_{r_jr_j}((t_1-t_2)+(g_1-g_3)) \gamma_{s_js_j}((t_1-t_2)+(g_2-g_4))| \\
+ & |\gamma_{r_js_j}((t_1-t_2)+(g_1-g_4))\gamma_{s_jr_j}((t_1-t_2)+(g_2-g_3))| \\
+ & \big| c_{r_js_jr_js_j}( t_1-t_2+g_1-g_4,t_1-t_2+g_2-g_4,g_3-g_4 ) \big| \Big\}\,.
\end{align*}
Evaluating the $ C_j(t_1,t_2) \, C_k(t_1,t_2) $ term in \eqref{lem_h1} leads to the consideration of terms which  are similar to  the following three:
\begin{align*}
I_{1,n}& =\frac{1}{16 \pi^4 b^4}\frac{1}{(n-b+1)^2} \sum_{t_1,t_2=1}^{n-b+1}\sum_{g_1,g_2,g_3,g_4=1}^b\sum_{v_1,v_2,v_3,v_4=1}^b\\
&  \big|\gamma_{r_jr_j}((t_1-t_2)+(g_1-g_3))\big|\big| \gamma_{s_js_j}((t_1-t_2)+(g_2-g_4))\big| \\
& \times \big|\gamma_{r_ks_k}((t_1-t_2)+(v_1-v_4))\big|\big|\gamma_{s_kr_k}((t_1-t_2)+(v_2-v_3))\big|,
\end{align*}
\begin{align*}
I_{2,n}& =\frac{1}{16 \pi^4 b^4}\frac{1}{(n-b+1)^2} \sum_{t_1,t_2=1}^{n-b+1}\sum_{g_1,g_2,g_3,g_4=1}^b\sum_{v_1,v_2,v_3,v_4=1}^b\\
&  \big|\gamma_{r_jr_j}((t_1-t_2)+(g_1-g_3))\big|\big| \gamma_{s_js_j}((t_1-t_2)+(g_2-g_4))\big| \\
& \times\big| c_{r_ks_kr_ks_k}( t_1-t_2+v_1-v_4,t_1-t_2+v_2-v_4,v_3-v_4 ) \big|
\end{align*}
and
\begin{align*}
I_{3,n}& =\frac{1}{16 \pi^4 b^4}\frac{1}{(n-b+1)^2} \sum_{t_1,t_2=1}^{n-b+1}\sum_{g_1,g_2,g_3,g_4=1}^b\sum_{v_1,v_2,v_3,v_4=1}^b\\
&  \big| c_{r_js_jr_js_j}( t_1-t_2+g_1-g_4,t_1-t_2+g_2-g_4,g_3-g_4 ) \big| \\
& \times\big| c_{r_ks_kr_ks_k}( t_1-t_2+v_1-v_4,t_1-t_2+v_2-v_4,v_3-v_4 ) \big|.
\end{align*}
For  $I_{1,n}$ we get summing first  over $v_2$, then over  $v_1$, then over $g_2$ and finally over $t_1$, that this term is $ {\mathcal O}(b/(n-b+1))$. For the term $ I_{2,n}$ we get by summing first over $ v_3$, then over $v_2$, then over $v_1$, then over $ g_2 $ and finally over $t_1$ that this term is ${\mathcal O}(1/(n-b+1))$. For $I_{3,n}$ we get summing first over $v_3$, then over $v_2$ , then over $v_1$, then over $g_3$, then over $ g_2 $ and finally over $t_1$, that  $I_{3,n}={\mathcal O}(1/(b(n-b+1)))$.  From this we conclude that
$ V_{1,n}(\ell_1,\ell_2) = {\mathcal O}(b/(n-b+1))$.

It remains to show that $ V_{3,n}(\ell_1,\ell_2) = {\mathcal O}(b/(n-b+1))$. For this  we get
\begin{align*}
\big|V_{3,n}&(\ell_1,\ell_2)\big|  \leq \frac{1}{16 \pi^4 b^4}\frac{1}{(n-b+1)^2} \sum_{t_1,t_2=1}^{n-b+1} \sum_{g_1,g_2, \cdots, g_8=1}^b\\
& \big|\textrm{cum}\big(X_{r_j}(t_1+g_1-1)\cdot X_{s_j}(t_1+g_2-1),  X_{r_k}(t_1+g_3-1)\cdot X_{s_k}(t_1+g_4-1),\\
& X_{r_j}(t_2+g_5-1)\cdot X_{s_j}(t_2+g_6-1), X_{r_k}(t_2+g_7-1)\cdot X_{s_k}(t_2+g_8-1)\big)\big|.
\end{align*}
The above cumulant term can be expressed as the sum of  products of cumulants over   all indecomposable partitions of the two dimensional table
\[ \begin{array}{cc}  t_1+g_1-1 & t_1+g_2-1 \\
 t_1+g_3-1 & t_1+g_4-1 \\  t_2+g_5-1 & t_2+g_6-1 \\  t_2+g_7-1 & t_2+g_8-1;
\end{array}\]
see  for instance  Brillinger (1981), Theorem 2.3.2.
Investigation of this sum shows that it   is dominated by those  indecomposable partitions which contain only pairs with a typical  term of such a partitions given by
\begin{align*}
 \frac{1}{16 \pi^4 b^4} & \frac{1}{(n-b+1)^2} \sum_{t_1,t_2=1}^{n-b+1} \sum_{g_1,g_2, \cdots, g_8=1}^b\\
& \Big|\textrm{cum}\big(X_{r_j}(t_1+g_1-1), X_{s_j}(t_2+g_6-1)\big)\\
& \times \textrm{cum}  \big(X_{r_k}(t_1+g_3-1), X_{s_k}(t_2+g_8-1)\big)\\
&\times \textrm{cum}\big(X_{s_j}(t_1+g_2-1), X_{r_j}(t_2+g_5-1)\big)\\
& \times \textrm{cum} \big(X_{s_k}(t_1+g_4-1), X_{r_k}(t_2+g_7-1)\big)\Big|\\
 = \frac{1}{16 \pi^4 b^4}  &\frac{1}{(n-b+1)^2} \sum_{t_1,t_2=1}^{n-b+1} \sum_{g_1,g_2, \cdots, g_8=1}^b\\
& \big|\gamma_{r_js_j}\big((t_1-t_2)+(g_1-g_6)\big)| \big|\gamma_{r_ks_k}\big((t_1-t_2)+(g_3-g_8)\big)|\\
 &\times \big|\gamma_{s_jr_j}\big((t_1-t_2)+(g_2-g_5)\big)| \big|\gamma_{s_kr_k}\big((t_1-t_2)+(g_4-g_7)\big)|.
\end{align*}
Summing first over $g_4$ , then over $g_2$, then over $g_3$ and then over $t_1$ we get that this term is $ {\mathcal O}(b/(n-b+1))$.
 $ \hfill \square $

\end{document}